\documentclass[journal, twoside, final]{IEEEtran}
\usepackage{amsmath,amssymb,amsfonts}
\usepackage{algorithm}
\usepackage{array}
\usepackage[caption=false,font=normalsize,labelfont={scriptsize,sf},textfont=sf]{subfig}
\usepackage{textcomp}
\usepackage{stfloats}
\usepackage{url}
\usepackage{graphicx}
\usepackage{cite}
\usepackage{xcolor}
\usepackage{algpseudocode}
\usepackage{float}
\usepackage{booktabs}
\usepackage{multirow}
\usepackage{longtable, threeparttable}
\usepackage{makecell}
\usepackage{siunitx}

\newcommand{\ABS}{\text{\rm ABS}}
\newcommand{\TBS}{\text{\rm TBS}}
\newcommand{\LoS}{\text{\rm L}}
\newcommand{\NLoS}{\text{\rm N}}

\newtheorem{theorem}{Theorem}
\newtheorem{proposition}{Proposition}

\newtheorem{lemma}{Lemma}
\newtheorem{remark}{Remark}

\begin{document}

\title{Vertical Heterogeneous Networks Beyond 5G: \\ CoMP Coverage Enhancement and Optimization}

\author{Tian Shi, Wenkun Wen,~\IEEEmembership{Member,~IEEE}, Peiran Wu,~\IEEEmembership{Member,~IEEE}, \\ and Minghua Xia,~\IEEEmembership{Senior Member,~IEEE}
 	\thanks{Received 27 July 2025; revised 29 September 2025; accepted 7 December 2025. The associate editor coordinating the review of this article and approving it for publication was F. Hou. \textit{(Corresponding author: Minghua Xia)}}
	\thanks{Tian Shi,  Peiran Wu, and Minghua Xia are with the School of Electronics and Information Technology, Sun Yat-sen University, Guangzhou 510006, China (email: shit26@mail2.sysu.edu.cn, wupr3@mail.sysu.edu.cn, xiamingh@mail.sysu.edu.cn).}
	\thanks{Wenkun Wen is with the R\&D Department of the Techphant Technologies Company Ltd., Guangzhou 510310, China (email: wenwenkun@techphant.net).}
	\thanks{Digital Object Identifier 10.1109/TWC.2025.3644244}
}

\markboth{IEEE Transactions on Wireless Communications} {Shi \MakeLowercase{\textit{et al.}}: Vertical Heterogeneous Networks Beyond 5G: CoMP Coverage Enhancement and Optimization}

\maketitle

\IEEEpubid{\begin{minipage}{\textwidth} \ \\[12pt] \centering 1536-1276 \copyright\ 2025 IEEE. All rights reserved, including rights for text and data mining, and training of artificial intelligence \\ and similar technologies. Personal use is permitted, but republication/redistribution requires IEEE permission. \\
See \url{https://www.ieee.org/publications/rights/index.html} for more information.\end{minipage}}

 \IEEEpubidadjcol

\maketitle

\begin{abstract}
Low-altitude wireless networks are increasingly vital for the low-altitude economy, enabling wireless coverage in high-mobility and hard-to-reach environments. However, providing reliable connectivity to sparsely distributed aerial users in dynamic three-dimensional (3D) spaces remains a significant challenge. This paper investigates downlink coverage enhancement in vertical heterogeneous networks (VHetNets) beyond 5G, where uncrewed aerial vehicles (UAVs) operate as emerging aerial base stations (ABSs) alongside legacy terrestrial base stations (TBSs). To improve coverage performance, we propose a coordinated multi-point (CoMP) transmission framework that enables joint transmission from ABSs and TBSs. This approach mitigates the limitations of non-uniform user distributions and enhances reliability for sparse aerial users. Two UAV deployment strategies are considered: \textit{i)} random UAV placement, analyzed using stochastic geometry to derive closed-form coverage expressions, and \textit{ii)} optimized UAV placement using a coverage-aware weighted $K$-means clustering algorithm to maximize cooperative coverage in underserved areas. Theoretical analyses and Monte Carlo simulations demonstrate that the proposed CoMP-enabled VHetNet significantly improves downlink coverage probability, particularly in scenarios with sparse aerial users. These findings highlight the potential of intelligent UAV coordination and geometry-aware deployment to enable robust, adaptive connectivity in low-altitude wireless networks.
\end{abstract}


\section{Introduction}
\label{sec:introduction}
\IEEEPARstart{L}{ow-altitude} wireless networks are emerging as a cornerstone of the low-altitude economy, enabled by technological innovation and increasingly diverse applications \cite{wu2025lowaltitude}. Among the key enablers are aerial platforms, particularly uncrewed aerial vehicles (UAVs), which function as modern \textit{aerial base stations} (ABSs) equipped with flexible deployment and high mobility, operating alongside legacy \textit{terrestrial base stations} (TBSs). Together, they form a vertical heterogeneous network (VHetNet) that integrates the advantages of both air- and ground-based infrastructures. This integrated architecture has gained significant attention for its potential to provide seamless communication services to both aerial and ground users. Thanks to their mobility and deployment flexibility, UAVs are well-suited to a range of scenarios, particularly during emergency response \cite{8123597}, disaster recovery \cite{7589913}, and large-scale events \cite{7744808}.

However, aerial users are often sparsely distributed, especially in dynamic three-dimensional (3D) airspace. In such settings, cooperative operation among multiple UAVs becomes essential to ensure reliable service. This paper addresses two deployment strategies for ABSs tailored to such scenarios: random deployment and intentional placement. In the former, stochastic geometry tools are employed to analytically evaluate downlink coverage under a coordinated multi-point (CoMP) transmission framework. In the latter, a coverage-aware weighted $K$-means clustering algorithm is proposed to optimize UAV placement, thereby improving coverage in underserved regions.

\subsection{Related Works and Motivation}
CoMP transmission is a well-established technique designed to enhance spectral efficiency and mitigate inter-cell interference \cite{6146494}. It has been extensively explored in heterogeneous networks, where cooperation among TBSs has been shown to improve both rate and reliability \cite{6928420}. Simulation studies \cite{5490976} and field trials \cite{5706317} have further validated the benefits of CoMP. To reduce the overhead associated with TBS searching, efficient implementations using Poisson-Delaunay triangulation have also been proposed for TBS cooperation \cite{8358978,9893878}.

Recently, CoMP strategies involving ABSs have gained significant attention. For instance, UAV swarms have been investigated for cooperative air-to-ground communications and formation control \cite{10381632}. Additionally, CoMP-based handoff schemes have been proposed to improve mobility and reliability in UAV-assisted networks beyond 5G \cite{10720695}. However, even with CoMP, aerial users may encounter coverage interruptions at certain altitudes due to altitude-dependent channel variability, which can degrade link quality.

 \IEEEpubidadjcol
 
VHetNets, which integrate ABSs and TBSs, have emerged as a promising solution for enhancing connectivity, coverage continuity, and link robustness. Previous studies have analyzed downlink performance using stochastic geometry under both line-of-sight (LoS) and non-line-of-sight (NLoS) conditions for ground users \cite{8833522}. Other research has explored altitude-aware association behavior and the benefits of ABSs in dense areas \cite{9250029}. Novel architectures, such as deploying ABSs on roadside infrastructure for optimized trajectory planning and enhanced aerial coverage, have also been proposed \cite{10225703}.

On a broader scale, stochastic geometry has been applied to satellite-integrated VHetNets to model uplink connectivity via terrestrial and aerial relays, considering terahertz propagation and association strategies \cite{10907788}. A spherical stochastic geometry framework has also been introduced to study global-scale VHetNet connectivity \cite{10438999}. However, transitioning from 2D to 3D space fundamentally alters the geometric structure—volume increases at a much faster rate than surface area, leading to increased data sparsity and reduced point density \cite{9151343}. These challenges complicate reliable connectivity and necessitate the development of new analytical tools and deployment strategies.

Stochastic geometry provides a robust analytical framework for modeling such networks. The Poisson point process (PPP) \cite{6042301} is commonly used for TBSs due to its mathematical tractability, while the binomial point process (BPP) \cite{haenggi2013stochastic,5299075} is better suited for modeling finite ABS deployments. BPP-based models have been applied to UAV networks under Nakagami-$m$ fading \cite{7967745}, as well as for joint uplink–downlink analysis in UAV networks \cite{10925893}. A key challenge in modeling CoMP in VHetNets lies in analytically deriving the joint distance distributions between a typical user and multiple cooperating ABSs and TBSs in 3D space. For brevity, Table~\ref{tab1} compares the relevant works.

The sparse distribution of aerial users, combined with the analytical tractability of stochastic geometry, highlights both the necessity and feasibility of developing CoMP strategies for VHetNets. Such strategies are essential for enhancing system capacity, improving user experience, and ensuring robust, seamless connectivity across 3D space.

\begin{table}
\begin{center}
\renewcommand\arraystretch{1.25}
\caption{Comparison of Related Works on VHetNets}
\label{tab1}
\begin{tabular}{!{\vrule width1.2pt} c !{\vrule width1.2pt} c | c | c | c !{\vrule width1.2pt}}
\Xhline{1.2pt}
\textbf{Ref.} & \textbf{ABSs} & \textbf{TBSs} &  \textbf{G2A Model} & \textbf{Tx Scheme} \\
\Xhline{1.2pt}
\cite{9893878} & N/A  & PPP & LoS link & CoMP\\
\hline
\cite{10381632,10720695} & PPP  & N/A & LoS link & CoMP\\
\hline
\cite{8833522} & PPP  & PPP & LoS/NLoS link & Single\\
\hline
\cite{10225703} & PLCP  & PPP & LoS/NLoS link & Single\\
\hline
\cite{10907788,10438999} & 3D PPP  & N/A & LoS link & Single\\
\hline
\cite{9151343} & 3D BPP  & N/A & LoS link & CoMP\\
\hline
\cite{7967745} & BPP  & N/A & LoS link & Single\\
\hline
\cite{8692749,8269068} & N/A  & PPP & LoS/NLoS link & Single\\
\hline
\cite{9250029,10925893,8644511} & BPP  & PPP & LoS/NLoS link & Single\\
\hline
This paper & BPP  & PPP & LoS/NLoS link & CoMP\\
\Xhline{1.2pt}
\end{tabular}
\end{center}
\end{table}

\subsection{Contributions}
This paper presents a comprehensive study on CoMP-enabled VHetNets, focusing on both analytical coverage analysis and UAV deployment optimization. The main contributions are as follows:
\begin{enumerate}
	\item \textit{Novel Network Model:} We propose a 3D VHetNet architecture that facilitates joint CoMP transmission among triads of ABSs and TBSs, using Poisson-Delaunay triangulation. This cooperative model significantly enhances service quality for spatially distributed aerial users while minimizing the overhead of cooperation.

	\item \textit{Analytical Distance Distributions:} In contrast to prior works that focus only on the nearest base station, we derive both marginal and joint probability density functions (PDFs) for distances to the general $n$-nearest ABSs and TBSs ($n \ge 1$), laying the foundation for user association and coverage evaluation. Notably, a two-regime user association behavior is identified.

	\item \textit{Deployment Optimization:} We examine two deployment strategies: for randomly deployed ABSs, we evaluate performance analytically using stochastic geometry; for optimal deployment, we propose a coverage-aware weighted $K$-means clustering algorithm to guide ABS placement toward coverage-deficient regions.
	
	\item \textit{Coverage Analysis:} We derive closed-form expressions for the downlink coverage probability of aerial users under CoMP transmission. Monte Carlo simulations confirm the analytical accuracy and demonstrate the substantial performance gains enabled by the proposed UAV-based CoMP strategies.
\end{enumerate}

 The remainder of the paper is organized as follows: Section~\ref{sec:system_model} describes the system and channel models. Section~\ref{Section-Distributions} characterizes the distributions of service distance and received signal.  Sections~\ref{sec:association_policy} and~\ref{sec:coverage_probability} analyze user association behavior and network coverage probability, respectively. Section~\ref{sec:deployment_optimization} presents an optimized UAV deployment strategy. Numerical results and discussions are given in Section~\ref{sec:numerical_results}, followed by conclusions in Section~\ref{sec:conclusions}.

\section{System and Channel Models} \label{sec:system_model}
We consider a VHetNet comprising legacy TBSs and emerging ABSs, with a focus on analyzing downlink performance under a CoMP transmission to sparse aerial user equipments (UEs).

\subsection{System Model}
\begin{figure}[!t]
  \centering
  \includegraphics[width=3.25in]{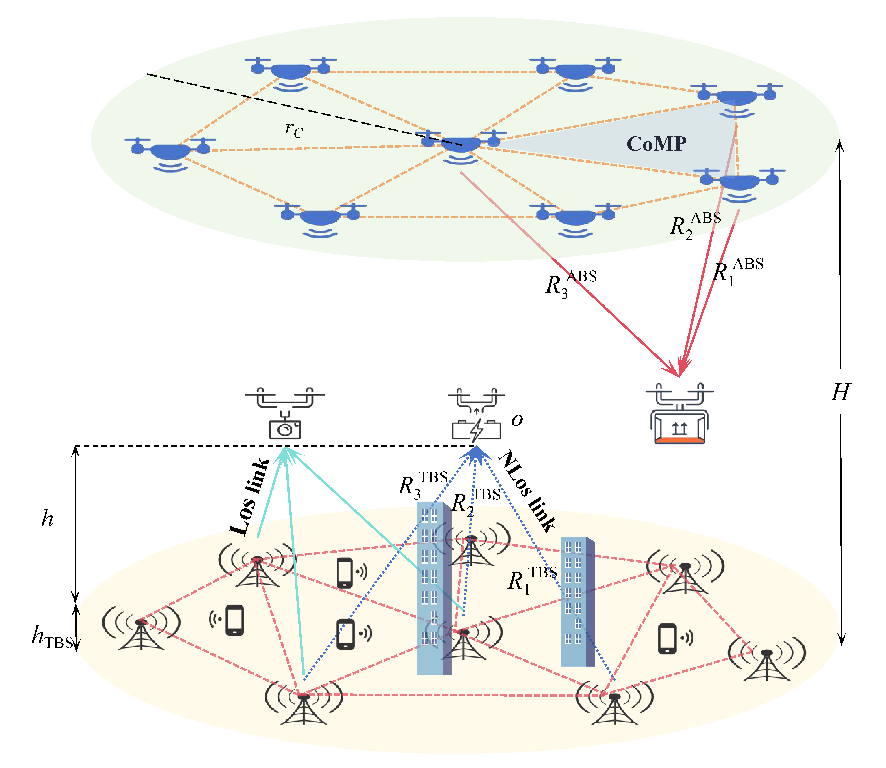}
  \vspace{-5pt}
  \caption{An illustration of a vertical heterogeneous network (VHetNet) comprising legacy terrestrial base stations (TBSs) and emerging aerial base stations (ABSs). The air-to-air (A2A) links between spatial UEs and ABSs typically experience line-of-sight (LoS) propagation due to their elevated positions. In contrast, the ground-to-air (G2A) links between TBSs and spatial UEs may undergo either LoS or non-line-of-sight (NLoS) propagation, depending on the environment and obstructions. To enhance network coverage performance, a coordinated multi-point (CoMP) transmission strategy is adopted in this work.}
  \label{fig:net}
  \end{figure}

As illustrated in Fig.~\ref{fig:net}, the aerial network comprises $N$ UAVs, each equipped with an ABS hovering at the same altitude of $H$. The locations of these ABSs are modeled as a finite BPP, denoted by $\Phi_{\ABS}$, uniformly distributed within a circular region of radius $r_C$, centered at $(0, 0, H)$. In contrast, the TBSs are assumed to follow a homogeneous PPP, denoted by $\Phi_{\TBS}$, with intensity $\lambda_{\TBS}$ and located at height $h_{\TBS}$. The overall network is then defined as the union $\Phi = \Phi_{\ABS} \cup \Phi_{\TBS}$, where $\Phi_{\ABS}$ and $\Phi_{\TBS}$ are assumed to be statistically independent, i.e., $\Phi_{\ABS} \perp \Phi_{\TBS}$.

To distinguish between aerial and terrestrial base station tiers, we define a tier indicator function $\kappa(x)$ for any node $x \in \Phi$:
\begin{equation*}
  \kappa(x)=
  \left\{\begin{array}{rl}
    \ABS, & \text{if } x\in\Phi_{\ABS}; \\
    \TBS, & \text{if } x\in\Phi_{\TBS}.
  \end{array} \right.
\end{equation*}
When an aerial UE communicates with a TBS, the wireless link may experience either LoS or NLoS propagation due to terrestrial obstructions, as depicted in Fig.~\ref{fig:net}. In contrast, the link between an aerial UE and an ABS is assumed always to be LoS. Let $x$ denote the location of a transmitter and define the link distance as $r = \| x \|$. We introduce an indicator variable $\zeta(x)\in\{\LoS,\NLoS\}$ to denote the propagation state of the link, where $\zeta(x) = \LoS$ if the link is LoS, and $\zeta(x) = \NLoS$ otherwise. Conditioned on $r$, the variable $\zeta(x)$ is modeled as a Bernoulli random variable whose success probability depends on the tier of the transmitter:
\begin{equation*}
  \mathbb P \bigl[\zeta(x)=\LoS \mid r\bigr]
  =
  \left\{\begin{array}{rl}
    1, & \text{if } \kappa(x)=\ABS; \\
    P_{\LoS}(r), & \text{if } \kappa(x)=\TBS,
  \end{array} \right.
\end{equation*}
where $0 \le P_{\LoS}(r) \le 1$ denotes the LoS probability for TBS links, and the corresponding NLoS probability is defined as $P_{\NLoS}(r) \triangleq 1 - P_{\LoS}(r)$.

For any parameter or random variable $X$ that depends on both the serving tier $\kappa \in \{\ABS, \TBS\}$ and the propagation condition $\zeta \in \{\LoS, \NLoS\}$, we adopt the unified notation:
\begin{equation*}
X_{\kappa,\zeta} =
  \left\{ \begin{array}{rl}
X_{\ABS},      & \text{if }  \kappa=\ABS; \\
X_{\TBS,\zeta},& \text{if }  \kappa=\TBS.
  \end{array} \right.
\end{equation*}
This convention extends similarly for superscripts, e.g., $X^{(\kappa, \zeta)}$. For instance, $\alpha_{\kappa, \zeta}$ corresponds to the path loss exponent, which simplifies to $\alpha_{\ABS}$ for ABS links, and distinguishes between $\alpha_{\TBS, \LoS}$ and $\alpha_{\TBS, \NLoS}$ for TBS links, depending on whether the propagation condition is LoS or NLoS.

Remarkably, our system model is primarily focused on providing downlink service to aerial users and does not explicitly account for terrestrial users. However, in the presence of co-channel terrestrial uplink activity, additional interference is introduced at ground level, leading to a decrease in the aerial signal-to-interference ratio (SIR). As the density or transmit power of these terrestrial uplink users increases, the coverage for aerial users degrades monotonically. In the PPP model, this additional interference is captured as an extra Laplace factor in the interference term, which results in a strictly lower success probability for aerial users.

\subsection{Channel Model}
In this study, we assume that aerial users hover at altitudes above surrounding buildings. Consequently, the air-to-air (A2A) links between a typical aerial UE and the ABSs are modeled as LoS channels, consistent with prior works~\cite{7414180, 9151343}. In contrast, existing ground-to-air (G2A) models have primarily been developed for {\it terrestrial users} located at approximately $1.5$ meters in height and, therefore, do not accurately reflect the propagation characteristics experienced by {\it aerial users} at higher altitudes. To address this, our model assumes TBSs are deployed at elevated positions, such as 30-meter macro base stations \cite{1543924}. Under these conditions, G2A links are subject to both LoS and NLoS propagation, as supported by empirical studies \cite{8644511}.

To model the LoS probability for G2A links, we adopt a simplified expression proposed in \cite{9250029}. Specifically, the LoS probability, denoted by $P_{\rm L}(z)$, is given by
\begin{equation} \label{eq:LoS_probability}
	P_{\rm L}(z) = -a\exp\left(-b\delta\right)+c,
\end{equation}
where $\delta = \arctan(h/z)$ is the elevation angle, $z$ represents the horizontal distance between the aerial user and the base station, and $a$, $b$, and $c$ are environment-specific parameters that implicitly capture the relative height difference between the aerial user and the TBS. For instance, $(a, b, c) = (1, 6.581, 1)$ in subarban areas and $(a, b, c) = (1.124, 0.049, 1.024)$ in highrise urban areas \cite{9250029}. 

Both A2A and G2A channels are assumed to undergo Nakagami-$m$ fading. The fading amplitude, denoted by ${\rm H}^{(\kappa,\zeta)}$, is modeled as a Nakagami random variable: ${\rm H}^{(\kappa,\zeta)}\sim \text{Nakagami}(m_{\kappa,\zeta}, \Omega)$, where $m_{\kappa,\zeta} \ge 0.5$ is the fading severity (shape) parameter and $\Omega = \mathbb{E}[( {\rm H}^{(\kappa,\zeta)})^{2}]$ denotes the average received power. The corresponding PDF is given by \cite{NAKAGAMI19603}:
\begin{equation}\label{eq:PDF_channel_gain}
  f_{{\rm H}^{(\kappa,\zeta)}}(x) = \frac{2m_{\kappa,\zeta}^{m_{\kappa,\zeta}} x^{2{m_{\kappa,\zeta}}-1}}{\Gamma(m_{\kappa,\zeta}) \, \Omega^{m_{\kappa,\zeta}}}  \exp\left(-\frac{{m_{\kappa,\zeta}}}{\Omega} x^2 \right), \, x \ge 0.
\end{equation}

Assuming all base stations transmit with equal power and that thermal noise is negligible compared to interference, the instantaneous SIR at a typical aerial UE can be expressed as
\begin{equation} \label{Eq-SIR}
  \Gamma_{\chi} \triangleq \frac{S_{\chi}}{I} = \frac{\left(\sum_{n\in \mathcal C_\chi}\left|{\rm H}^{(\kappa_n,\zeta_n)}_{n}\right|(R_n)^{-\alpha_{\kappa_n,\zeta_n}/2}\right)^2}{\sum_{\Upsilon\in Q}\sum_{k\in \Phi_{\Upsilon} \backslash \mathcal C_\chi}{\rm G}_k^{(\Upsilon,\zeta_k)}(D_k)^{-\alpha_{\Upsilon,\zeta_k}}}, \, \chi \in Q,
\end{equation}
where $R_n$ and $D_k$ denote the distances from a typical UE to the $n^\text{th}$ cooperating base station and the $k^\text{th}$ interferer, respectively; $\alpha_{\kappa, \zeta}$ is the path loss exponent associated with tier $\kappa \in \{\ABS, \TBS\}$ and link state $\zeta \in \{\LoS, \NLoS\}$;
${\rm H}^{(\kappa,\zeta)}$ and ${\rm G}^{(\kappa,\zeta)} = ({\rm H}^{(\kappa,\zeta)})^2$ represent the fading amplitude and channel power gain, respectively;
$S_{\chi}$ denotes the aggregated received signal power from cooperating base stations in the coordination set $\mathcal C_\chi$;
$I$ is the total interference from all other non-cooperating base stations, and $Q$ is the set of all base station tiers.

Notably, the received power ${\rm G}^{(\kappa,\zeta)}$ follows a Gamma distribution with shape $m_{\kappa,\zeta}$ and scale $\Omega/m_{\kappa,\zeta}$, i.e., ${\rm G}^{(\kappa,\zeta)} \sim \Gamma(m_{\kappa,\zeta}, \Omega/m_{\kappa,\zeta})$. For convenience, the key notations used throughout this paper are summarized in Table~\ref{tab:notation}.

\begin{table}[!t]
\renewcommand\arraystretch{1.05}
\begin{center}
\caption{Summary of Notation}
\centering
\scriptsize
\label{tab:notation}
\begin{tabular}{!{\vrule width1.2pt} c !{\vrule width1.2pt} c !{\vrule width1.2pt}}
\Xhline{1.2pt}
\textbf{Notation} & \textbf{Description}\\
\Xhline{1.2pt}
\multirow{2}{*}{$h$} & Height of the aerial user with reference \\
  &  to the TBSs \\
\hline
\multirow{2}{*}{$H$} & Height of the ABSs with reference \\
  &  to the TBSs\\
\hline
\multirow{2}{*}{$r_C$} & Radius of the circle where ABSs \\
  & are distributed \\
\hline
\multirow{2}{*}{$\kappa_n,\zeta_n$} & The tier label and the LoS/NLoS state \\
  &  of the $n^{\text{\rm th}}$ link\\
\hline
\multirow{2}{*}{$R_n^{\ABS},R_n^{\TBS}$} & Distance between the aerial user \\
  & and its $n^{\text{th}}$ nearest ABS or TBS, respectively\\
\hline
\multirow{2}{*}{$D_k^{\ABS},D_k^{\TBS}$} & Distance between the aerial user \\
  & and interfering ABS or LoS TBS, respectively\\
\hline
\multirow{2}{*}{$P_{\LoS}(\cdot),P_{\NLoS}(\cdot)$} & Probability of the aerial user being \\
  &in LoS and NLoS with TBS, respectively \\
\hline
\multirow{2}{*}{$m_{\ABS},m_{\TBS,\LoS},m_{\TBS,\NLoS}$} & Nakagami-$m$ fading parameter for ABS, \\
& LoS TBS, or NLoS TBS, respectively\\
\hline
\multirow{2}{*}{$\alpha_{\ABS},\alpha_{\TBS,\LoS},\alpha_{\TBS,\NLoS}$} & Path loss parameter for ABS, LoS TBS, \\
&or NLoS TBS, respectively\\
\hline
$\lambda_{\TBS}$ & Density of terrestrial base stations \\
\hline
\multirow{2}{*}{$\Phi_{\ABS}$,$\Phi_{\TBS}$,$\Phi$} & Tier of the ABSs, the TBSs, \\
&or the overall BSs, respectively \\
\hline
\multirow{3}{*}{$\mathrm{H}_n^{\ABS},\mathrm{H}_n^{(\TBS,\LoS)},\mathrm{H}_n^{(\TBS,\NLoS)}$} & Channel fading amplitude of the $n^{\rm{th}}$ link  \\
&between the aerial user and an ABS, LoS TBS, \\
&or NLoS TBS, respectively \\
\hline
\multirow{3}{*}{$\mathrm{G}_k^{\ABS},\mathrm{G}_k^{(\TBS,\LoS)},\mathrm{G}_k^{(\TBS,\NLoS)}$} & Channel power gain of the $k^{\rm{th}}$ link\\
& between the aerial user and an ABS, LoS TBS, \\
& or NLoS TBS, respectively \\
\hline
 \multirow{2}{*}{$\mathcal{A}_{\ABS},\mathcal{A}_{\TBS}$} & Probability that the typical user is associated  \\
 &with three ABSs or TBSs  \\
\hline
\multirow{2}{*}{$I,\mathcal{L}_{\sqrt{I}},\mathcal{L}_{I}$} & Interference, Laplace transform of $\sqrt{I}$, \\
&and Laplace transform of interference \\
\hline
\multirow{2}{*}{$P_{\ABS},P_{\TBS}$}  & Coverage probability  given that the typical user \\
& is associated with three ABSs or TBSs \\
\Xhline{1.2pt}
\end{tabular}
\end{center}
\vspace{-15pt}
\end{table}

\section{Mathematical Preliminary: Distributions of Service Distance and Received Signal}
\label{Section-Distributions}

Due to the sparse and spatially dispersed distribution of aerial UEs in 3D space, ensuring reliable coverage often requires coordinated transmission from multiple base stations. However, to the best of our knowledge, the statistical characterization of service distances—specifically, the distribution of the distance to the $n^\text{th}$-nearest base station, and the joint distribution of distances to the nearest $n > 1$ base stations—has not been explicitly derived for 3D VHetNets. Such distributions are essential for accurately analyzing the performance of CoMP strategies and form the mathematical foundation for the subsequent analysis in this paper. Moreover, these results offer analytical tools for broader research on cooperative aerial networks.

\subsection{Distributions of Service Distance}

We first characterize the service distance from a typical aerial UE to the $n^\text{th}$-nearest ABS.
\begin{lemma}[ABS Case] \label{lem:A2A_PDF_n_distance}
  The PDF of the distance from a typical aerial UE to the $n^\text{\rm th}$ nearest ABS is given by
  \begin{align} \label{eq:A2A_PDF_n_distance}
    f_{R_{n}}^{\ABS}(r)
    &=\frac{N!}{(n - 1)!(N - n)!}\left(\frac{2r}{r_C^2}\right)\left(\frac{r^2 - (H-h)^2}{r_C^2}\right)^{n - 1} \nonumber \\
    &\quad\times\left(\frac{r_{\max}^2 - r^2}{r_C^2}\right)^{N - n},
  \end{align}
  where $H-h\leq r\leq r_{\max}$, with $r_{\mathrm{max}} \triangleq \sqrt{(H-h)^2+r_C^2}$. The joint PDF of the distances to the nearest $n > 1$ ABSs is given by
  \begin{align} \label{eq:A2A_JPDF_distance}
    \lefteqn{f_{R_1, R_2,\cdots, R_n}^{\ABS}(r_1,r_2,\cdots,r_n)} \nonumber \\
    &=\frac{N!}{(N-n)!} \left(\frac{2}{r_C^2} \right)^n r_1r_2\cdots r_n \left(\frac{r_{\max}^2 - r_n^2}{r_C^2}\right)^{N - n},
  \end{align}
  where $H-h\leq r_1\leq r_2\leq \cdots\leq r_n\leq r_{\max}$.
\end{lemma}

\begin{IEEEproof}
  See Appendix \ref{sec:proof_A2A_PDF_n_distance}.
\end{IEEEproof}

For the special case of $n = 1$, \eqref{eq:A2A_PDF_n_distance} simplifies to
\begin{align} \label{eq:A2A_PDF_n_distance_reduced}
    f_{R_{1}}^{\ABS}(r)
    &=N\left(\frac{2r}{r_C^2}\right)\left(\frac{r_{\max}^2 - r^2}{r_C^2}\right)^{N - 1},
\end{align}
which matches the result reported in \cite[Eq. (7)]{7967745}.

Next, we derive the corresponding distribution for TBSs.
\begin{lemma}[TBS Case] \label{lem:G2A_PDF_n_distance}
  The PDF of the distance from a typical aerial UE to the $n^\text{\rm th}$ nearest TBS is given by
  \begin{align}
    f_{R_{n}}^{\TBS}(r)
    &=\frac{2(\pi\lambda_{\TBS})^n r}{\Gamma(n)} \left(r^{2}-h^{2}\right)^{n-1} \nonumber \\
    &\quad \times\exp\left(-\pi\lambda_\TBS(r^{2}-h^{2})\right),
   \label{eq:G2AL_PDF_n_distance}
  \end{align}
  where  $r > h$. The joint PDF of the distances to the $n > 1$ nearest TBSs is given by
  \begin{align}
    &f_{R_1, R_2,\cdots, R_n}^{\TBS}(r_1,r_2,\cdots,r_n) \nonumber  \\
    &=(2\pi\lambda_{\TBS})^n\prod_{i = 1}^{n}r_i\exp\left(-\pi\lambda_{\TBS}(r_n^{2}-h^{2})\right),
    \label{eq:G2AL_JPDF_distance}
   \end{align}
   where $h\leq r_1\leq r_2 \leq \cdots \leq r_n$.
\end{lemma}

\begin{IEEEproof}
  See Appendix \ref{sec:proof_G2A_PDF_n_distance}.
\end{IEEEproof}

For the special case of $n = 1$, \eqref{eq:G2AL_PDF_n_distance} reduces to
\begin{align}
    f_{R_{1}}^{\TBS}(r)&=2\pi\lambda_{\TBS} r\exp\left(-\pi\lambda_\TBS (r^{2}-h^{2})\right), \label{eq:G2AL_PDF_n_distance-reduced}
\end{align}
which is consistent with \cite[Eq. (6)]{8644511} under the assumption $P_\textnormal{L}(\cdot) = 1$.

While the derived service distance distributions enable accurate performance evaluation, increasing the cooperation size $n$ enhances coverage but also introduces additional coordination overhead. As shown in the literature \cite{8358978, 9893878, 9644611, 10288566}, cooperation among three base stations significantly improves coverage probability compared to single- or dual-base station CoMP. However, the marginal gain from adding a fourth or more base stations becomes negligible. Therefore, an effective trade-off is achieved when each aerial UE is served by exactly three base stations, forming the CoMP set $\mathcal{C}_\chi$ with cardinality $|\mathcal{C}_\chi| = 3$. Specifically,
\begin{align*}
  \mathcal C_\chi
  &= \{x_1,x_2,x_3\}\subset\Phi_\chi,\quad  \chi \in Q \triangleq \{\ABS,\,\TBS\},\\
  \lVert x_1\rVert &\le \lVert x_2\rVert \le \lVert x_3\rVert
  \le \lVert x\rVert,\quad \forall\,x\in\Phi_\chi\setminus\mathcal C_\chi.
\end{align*}
For each $n=1, 2, 3$, we define the link distance $r_n = \lVert x_n\rVert$ and the propagation condition $\zeta_n = \zeta(x_n)$.

\begin{figure}[!t]
  \centering
  \includegraphics[width=3.25in]{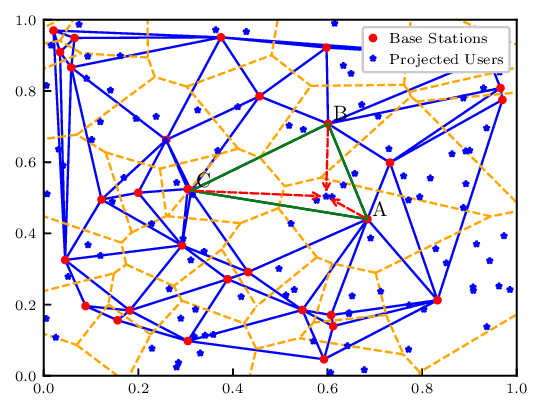}
  \vspace{-10pt}
  \caption{An illustrative Poisson-Delaunay triangulation network where each CoMP set consists of either TBSs or ABSs located at the vertices of Delaunay triangles  (outlined by blue solid edges). Red solid dots represent the positions of TBSs or ABSs. Blue stars denote aerial users, projected onto the 2D plane for association purposes. ABSs are distributed according to a finite Binomial Point Process (BPP) within a circular region of radius $r_\text{C}$. At the same time, TBSs follow a homogeneous Poisson Point Process (PPP) over a normalized area of 1~km\textsuperscript{2}.}
  \label{fig:Delaunay}
\end{figure}

In the presence of both ABSs and TBSs, if each aerial user associates with its closest base station, the resulting spatial partitioning follows a \textit{Poisson–Voronoi tessellation}, as is typical in stochastic geometry. However, the mathematical treatment of such tessellations is challenging due to their structural complexity \cite{8358978}. To address this, we consider the \textit{Poisson–Delaunay triangulation}, which is the dual of the Voronoi diagram and offers more tractable analytical properties \cite{8358978}. The Delaunay triangulation can be constructed using standard algorithms, e.g., Fortune’s sweep-line method, Bowyer–Watson insertion, or divide-and-conquer techniques \cite{Hjelle2006}.

Fig.~\ref{fig:Delaunay} illustrates the 2D projection of the 3D scenario shown in Fig.~\ref{fig:net}, considering either the TBS or ABS tier. Red dots indicate base station locations, blue stars represent projected aerial users, and the yellow dashed lines mark the Poisson–Voronoi cell boundaries. The blue solid lines correspond to the edges of the dual Delaunay triangulation. Each triangle in the Delaunay graph defines a CoMP cluster composed of three base stations, such as the cluster $\mathcal C_\chi = \{A, B, C\}$, which cooperatively serve users located within the interior of the triangle. Without loss of generality, we assume a typical aerial user is located at $(0, 0, h)$, where $0 \leq h \leq H$, and is served by its three nearest base stations, which may belong to either the TBS set $\Phi_\TBS$ or the ABS set $\Phi_\ABS$.

\subsection{Distributions of Received Signal}
As illustrated in Fig.~\ref{fig:net}, a CoMP set comprising either three ABSs or three TBSs forms a Delaunay triangle to cooperatively serve a typical aerial user. Let $\boldsymbol{\zeta} = (\zeta_1, \zeta_2, \zeta_3) \in \{\LoS,\NLoS\}^3$ denote the LoS or NLoS state of each serving base station (BS) in the CoMP set $\mathcal{C}_\chi$.

To facilitate the derivation of user association and coverage probability in subsequent sections, we define the aggregated received signal strength as
\begin{equation}
	U_{\chi,\boldsymbol{\zeta}} \triangleq \sum_{n\in \mathcal C_\chi}\left|{\rm H}^{(\chi,\zeta_n)}_{n}\right|(R_n)^{-\alpha_{\chi,\zeta_n}/2},
\end{equation}
and the corresponding large-scale fading component as
\begin{equation}
V_{\chi,\boldsymbol{\zeta}} \triangleq \sum_{n\in \mathcal C_\chi} (R_n)^{-\alpha_{\chi,\zeta_n}/2}.
\end{equation}

To enable tractable analysis, the distributions of $U_{\chi,\boldsymbol{\zeta}}$ and $V_{\chi,\boldsymbol{\zeta}}$ are approximated using a Gamma distribution, as detailed in the following lemma.
\begin{subequations}\label{eq:U_V}
\begin{lemma}\label{lem:PDF_T}
The PDFs of $U_{\chi,\boldsymbol{\zeta}}$ and $V_{\chi,\boldsymbol{\zeta}}$ can be approximated by Gamma distributions as
\begin{align}
  f_{U_{\chi,\boldsymbol{\zeta}}}(x) &= \frac{x^{\nu_{\chi,\boldsymbol{\zeta}} - 1}}{\theta_{\chi,\boldsymbol{\zeta}}^{\nu_{\chi,\boldsymbol{\zeta}}} \Gamma(\nu_{\chi,\boldsymbol{\zeta}})}\exp\left(-\frac{x}{\theta_{\chi,\boldsymbol{\zeta}}}\right), \label{eq:PDF_T} \\
  f_{V_{\chi,\boldsymbol{\zeta}}}(x) &= \frac{x^{\nu_{\chi,\boldsymbol{\zeta}}' - 1}}{\left(\theta_{\chi,\boldsymbol{\zeta}}'\right)^{\nu_{\chi,\boldsymbol{\zeta}}'} \Gamma(\nu_{\chi,\boldsymbol{\zeta}}')} \exp\left(-\frac{x}{\theta_{\chi,\boldsymbol{\zeta}}'}\right), \label{eq:PDF_T_p}
\end{align}
where the shape and scale parameters are given by
  \begin{align}
    \nu_{\chi,\boldsymbol{\zeta}} &= \frac{(\sum_{n=1}^{3}A_n\Delta_n)^2}{\Omega\sum_{n=1}^{3}B_n+\sum_{p\neq q}^{3}\Delta_p \Delta_q C_{p,q}  - (\sum_{n=1}^{3}A_n\Delta_n)^2}, \nonumber \\
    \theta_{\chi,\boldsymbol{\zeta}} &= \frac{\Omega\sum_{n=1}^{3}B_n+\sum_{p\neq q}^{3}\Delta_p \Delta_q C_{p,q}  - (\sum_{n=1}^{3}A_n\Delta_n)^2}{\sum_{n=1}^{3}A_n\Delta_n}, \nonumber \\
    \nu_{\chi,\boldsymbol{\zeta}}' &= \frac{(\sum_{n=1}^{3} A_n)^2}{\sum_{n=1}^{3}B_n+\sum_{p\neq q}^{3}C_{p,q}-(\sum_{n=1}^{3}A_n)^2}, \label{eq:nu_p} \\
    \theta_{\chi,\boldsymbol{\zeta}}' &= \frac{{\sum_{n=1}^{3}B_n+\sum_{p\neq q}^{3}C_{p,q}-(\sum_{n=1}^{3}A_n)^2}}{\sum_{n=1}^{3}A_n}. \label{eq:theta_p}
  \end{align}
The intermediate terms used above are defined as follows:
  \begin{align}
    A_n &= \int_{a_\chi}^{b_\chi}\left( r^{-\alpha_{\chi,\zeta_n}/2}P_{\zeta_n}(r) \right) f_{R_n}^\chi(r) \, \mathrm{d}r, \\
    B_n &=  \int_{a_\chi}^{b_\chi} (r^{-\alpha_{\chi,\zeta_n}}P_{\zeta_n}(r) f_{R_n}^\chi(r) \, \mathrm{d}r, \\
    C_{p,q} &= 2\int\limits_{\substack{a_\chi\leq r_1 \leq r_2 \leq r_3\leq b_\chi}} \bigg(\sum_{\zeta_p,\zeta_q\in Q} P_{\zeta_p}P_{\zeta_q} \\
    &\quad r_p^{-\alpha_{\chi,\zeta_q}/2} r_q^{-\alpha_{\chi,\zeta_p}/2} \bigg)\times f_{R_1, R_2, R_3}^\chi\,\mathrm{d}\boldsymbol{r}, \\
    \Delta_n&= \frac{\Gamma(m_{\chi,\zeta_n} + \frac{1}{2})}{\Gamma(m_{\chi,\zeta_n})} \left(\frac{\Omega}{m_{\chi,\zeta_n}}\right)^{\frac{1}{2}}
  \end{align}
with the integration bounds $(a_\chi, b_\chi)$ defined as
\begin{equation}
    (a_\chi, b_\chi) =
    \left\{\begin{array}{rl}
    (H-h, \, r_{\max}), & \text{if } \chi = \ABS; \\
    (h, \, \infty),  & \text{if }  \chi = \TBS.
    \end{array} \right.
  \end{equation}
\end{lemma}
\end{subequations}

\begin{IEEEproof}
  See Appendix \ref{sec:proof_PDF_T}.
\end{IEEEproof}

\section{Two-Regime User Association Behavior}
\label{sec:association_policy}
Before analyzing the network coverage probability, we first derive the association probability. In general, a user is served by three base stations, which can be all ABSs, all TBSs, or a combination of both. Association is determined based on the long-term average received power, conditioned on the link being LoS or NLoS. A Monte Carlo simulation was conducted to model a network scenario with $50,000$ users uniformly distributed within a 3D rectangular prism of dimensions $1000~\si{m} \times 1000~\si{m} \times (30 - 300)~\si{m} $. The simulation considered $30$ ABSs deployed within a circular area of radius \SI{500}{m}, and TBSs were modeled with a density of \SI{20}{\text{TBSs}/\kilo\meter\squared}. The TBS and ABS altitudes were fixed at \SI{30}{m} and \SI{320}{m}, respectively. The path loss exponents were set to $\alpha_{\ABS} = \alpha_{\TBS,\LoS} = 2$ and $\alpha_{\TBS,\NLoS} = 2.7$. At each aerial user altitude, the proportions of users associated with three base stations were calculated for each of the following configurations: all ABSs, all TBSs, or a mix of both.

The results shown in Fig.~3 reveal that mixed configurations are rare, accounting for less than $10\%$ of the cases and approximately $5\%$ on average. Based on these observations, we limit our analysis to two mutually exclusive association scenarios: users are served by either {\it i)} three ABSs (A2A scenario) or {\it ii)} three TBSs (G2A scenario). Let $\mathcal{A}\chi = \mathbb{P}(\mathcal{E}\chi)$ denote the probability of event $\mathcal{E}_\chi $, where $ \chi \in \{\ABS, \TBS\}$. In the A2A case, the user is associated with three ABSs, all of which have LoS links. In the G2A case, the three TBS links may exhibit any combination of LoS or NLoS conditions, resulting in $8$ possible link state combinations.

\begin{figure}[!t]
	\centering
	\includegraphics[width=3.25in]{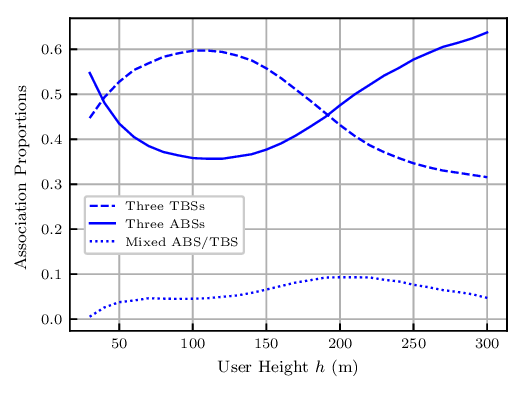}
	\vspace{-10pt}
	\caption{Association proportions of the three BS configurations (all ABS, all TBS, mixed) versus aerial user altitude $h$.}
	\label{fig:Configuration}
\end{figure}

\begin{remark}[Practical Considerations for Adopting the Same-Layer Three-Site CoMP Strategy]
  We adopt the \emph{same-layer three-site CoMP strategy} as the baseline for our analysis. In addition to the observations described above, this choice is motivated by several practical considerations: First, phase-level synchronization and low-latency, deterministic backhaul are more readily achievable within the same operator/vendor domain, with a uniform frame configuration across all base stations. In contrast, \emph{cross-tier ABS--TBS CoMP} faces challenges such as ABS clock drift in moving or semi-mobile scenarios, heterogeneous antenna and beam designs, and the complexities of combining air-to-ground and terrestrial backhaul with more stringent delay/jitter requirements. As a result, cross-tier cooperation is more commonly implemented as CS/CB-CoMP, while coherent CoMP is typically confined to same-tier clusters due to these operational challenges \cite{10288566}. Thus, we treat cross-tier CoMP as a high-cost, less common extension.
\end{remark}

Based on the same-layer three-site CoMP strategy, the probability that a typical UE associates with three ABSs is given in the following lemma.
\begin{lemma}\label{lem:association_probabilities}
  The probabilities that a typical user is associated with three ABSs are given by
  \begin{align}
    \mathcal{A}_{\ABS}&=\prod_{\boldsymbol{\zeta}\in \{\LoS,\NLoS\}^3}\int\limits_{\substack{H - h \leq r_1 \leq r_2 \\ r_2 \leq r_3 \leq r_{\max}}} \hspace{-1.6em} \frac{\gamma\left(\nu_{\TBS,\boldsymbol{\zeta}}',\left(\sum_{n=1}^3r_n^{-\frac{\alpha_{\ABS}}{2}}\right)/\theta_{\TBS,\boldsymbol{\zeta}}'\right)}{\Gamma(\nu_{\TBS,\boldsymbol{\zeta}}')}  \nonumber \\
  &\hspace{8em} \times f_{R_1, R_2, R_3}^{\ABS}(r_1,r_2,r_3) \, \mathrm{d}\boldsymbol{r},
 \label{eq:probability_A_ABS}
   \end{align}
   where $\gamma(\cdot, \cdot)$ denotes the lower incomplete Gamma function.
\end{lemma}

\begin{IEEEproof}
The association probability with ABSs is given by
\begin{align}
\mathcal{A}_{\ABS} &= \mathbb{P}\left(V_{\ABS} > V_{\TBS, \boldsymbol{\zeta}}, \ \forall \boldsymbol{\zeta} \in \{\LoS, \NLoS\}^3\right) \nonumber \\
&= \prod_{\boldsymbol{\zeta} \in \{\LoS, \NLoS\}^3} \mathbb{P}\left(V_{\ABS} > V_{\TBS,\boldsymbol{\zeta}}\right). \label{eq:association}
\end{align}

Applying the law of total probability over the distance vectors to ABSs yields:
\begin{align}
\mathcal{A}_{\ABS}
  &=\prod_{\boldsymbol{\zeta}\in \{\LoS,\NLoS\}^3}\int\limits_{\substack{H - h \leq r_1 \leq r_2 \\ r_2 \leq r_3 \leq r_{\max}}} \mathbb{P}\left(V_{\TBS,\boldsymbol{\zeta}}<\sum_{n\in \mathcal{C}_{\ABS}}r_n^{-\alpha_{\ABS}/2}\right) \nonumber \\
  &\hspace{8em}  \times f_{R_1, R_2, R_3}^{\ABS}(r_1,r_2,r_3)\mathrm{d} \boldsymbol{r}  \label{eq:association:law_of_total}
\end{align}
which, by evaluating the CDF of the Gamma-distributed variable $V_{\TBS,\boldsymbol{\zeta}}$, becomes the desired \eqref{eq:probability_A_ABS}.
\end{IEEEproof}

Using the complement rule, the TBS association probability is given by
\begin{equation}\label{eq:probability_A_TBS}
\mathcal{A}_{\TBS}=1-\mathcal{A}_{\ABS}.
\end{equation}

\begin{figure}[!t]
	\centering
	\includegraphics[width=3.25in]{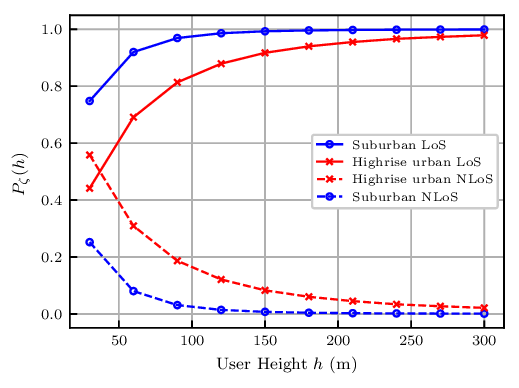}
	\vspace{-10pt}
	\caption{LoS ($\zeta = \LoS$) and NLoS ($\zeta = \NLoS$) probabilities of the G2A channel versus aerial user altitude $h$ in suburban and high-rise urban environments.} \label{fig:LoS_NLoS}
\end{figure}

\begin{figure}[!t]
	\centering
	\includegraphics[width=3.25in]{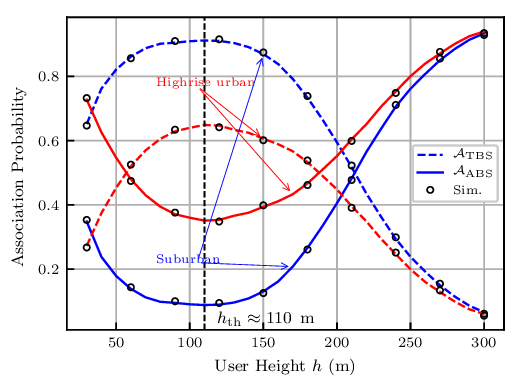}
	\vspace{-10pt}
	\caption{Association probabilities of an aerial user with the ABS and TBS versus aerial user altitude $h$, with ABS fixed at $H = 320$ \si{m}.}
	\label{fig:Association}
\end{figure}

Fig.~\ref{fig:LoS_NLoS} illustrates the LoS and NLoS probabilities of the G2A channel as a function of user altitude $h$ in suburban and dense high-rise urban environments. These propagation conditions significantly impact association behavior, as shown in Fig.~\ref{fig:Association}.

At low altitudes ($h = 30$ \si{m}), the NLoS probability exceeds 0.5 in the urban scenario and remains around 0.25 in the suburban one. Accordingly, the ABS maintains a dominant association probability of approximately 0.74 in the urban case but only 0.38 in the suburban one. As the user ascends ($30~\si{m} < h \leq 110~\si{m}$), the NLoS probability decreases, thereby improving the G2A link quality. Consequently, the TBS association probability peaks at 0.92 (suburban) and 0.65 (urban) around $h = 110$ \si{m}, while the ABS probability correspondingly declines.

For $h > 110$ \si{m}, the G2A channel becomes predominantly LoS (LoS $> 0.86$ in both environments). Owing to geometric proximity advantages, the ABS increasingly dominates the association, with its probability reaching 0.95 at $h = 300$ \si{m}, while the TBS probability falls below 0.05.

These observations reveal a clear {\it two-regime behavior}: at lower altitudes, {\it link blockage} is the dominant factor affecting association, whereas at higher altitudes, {\it geometric path loss} becomes the key determinant. This insight can guide the optimal deployment of ABSs and the design of association biasing in diverse urban morphologies, as formalized in the following proposition.
\begin{proposition}[Two-Regime Association Behavior]
\label{proposition:two_regimes}
The association behavior of aerial users exhibits a two-regime structure:
\begin{equation*}
	\mathcal{A}_{\ABS}(h) \left\{
	\begin{array}{ll}
		\text{increases with } h, & \text{if } h > h_{\text{\rm th}}; \\
		\text{decreases with } h, & \text{if } h < h_{\text{\rm th}},
	\end{array}
	\right.
\end{equation*}
where $h_{\text{\rm th}}$ is a critical threshold altitude (e.g., $h_{\text{\rm th}} \approx 110$ \si{m}) that separates two distinct regimes:
\begin{itemize}
	\item \textbf{Low-altitude regime ($h < h_{\text{\rm th}}$)}: LoS blockage dominates, favoring TBS association due to reduced signal obstruction.
	\item \textbf{High-altitude regime ($h > h_{\text{\rm th}}$)}: LoS conditions dominate, and geometric proximity to ABSs leads to stronger signal reception and increased ABS association.
\end{itemize}
 This threshold-based insight supports the design of altitude-aware association policies and optimized ABS management strategies in heterogeneous urban environments.
\end{proposition}

Building on the two-regime behavior identified in Proposition~\ref{proposition:two_regimes}, we now examine the specific altitudes where the ABS association probability reaches exactly $0.5$. The next proposition formalizes these heights, providing further insights into the dynamics of association at critical altitudes.

\begin{proposition}[0.5-Association Height]
\label{proposition:0.5_association_height}
Let $\mathcal{A}_{\ABS}:[H_{\min},H_{\max}]\to[0,1]$ be continuous and strictly U-shaped with unique minimizer $h_{\mathrm{th}}\in(H_{\min},H_{\max})$. Set
$m:=\mathcal{A}_{\ABS}(h_{\mathrm{th}}),
a_-:=\mathcal{A}_{\ABS}(H_{\min}),
a_+:=\mathcal{A}_{\ABS}(H_{\max}),$
and $\mathcal{H}_{0.5}:=\{h:\mathcal{A}_{\ABS}(h)=0.5\}$.
\begin{enumerate}
\item If $m>0.5$, then $\mathcal{H}_{0.5}=\emptyset$.
\item If $m=0.5$, then $\mathcal{H}_{0.5}=\{h_{\mathrm{th}}\}$.
\item If $m<0.5$, then
\begin{equation*}
  |\mathcal{H}_{0.5}|=\mathbf{1}\{a_-\ge0.5\}+\mathbf{1}\{a_+\ge0.5\}.
\end{equation*}
Moreover, each contributing endpoint determines one solution on its side:
\begin{itemize}
\item if $a_-\!>\!0.5$ (resp. $=\!0.5$), there is one solution in $(H_{\min},h_{\mathrm{th}})$ (resp. at $H_{\min}$);
\item if $a_+\!>\!0.5$ (resp. $=\!0.5$), there is one solution in $(h_{\mathrm{th}},H_{\max})$ (resp. at $H_{\max}$).
\end{itemize}
\end{enumerate}
\end{proposition}

Physically, when two solutions to $\mathcal{A}_{\ABS}(h)=0.5$ exist, the larger root $ h_{0.5}^{+} \in (h_{\text{\rm th}}, H_{\max}] $ can be interpreted as the \emph{handover-neutral height}. At this altitude, ABS and TBS association probabilities are balanced in the regime where $\mathcal{A}_{\ABS}(h)$ is increasing with $ h $, meaning that ABS dominance will continue to strengthen. Therefore, $ h_{0.5}^{+} $ is of particular importance for \emph{handover management}: it serves as a natural reference point for setting hysteresis margins to mitigate ping-pong handovers and for designing altitude-aware biasing strategies that adapt BS selection to user height.

In contrast, the smaller root $ h_{0.5}^{-} \in (H_{\min}, h_{\text{\rm th}}) $ lies in the regime where $\mathcal{A}_{\ABS}(h)$ is still decreasing. At this altitude, users are prone to frequent switching between ABS and TBS, as the association preference lacks stability. Hence, $ h_{0.5}^{-} $ represents an \emph{unfavorable operating point}, highlighting the necessity of avoiding coverage configurations that force users to hover around this altitude region.

\section{Network Coverage Probability Analysis} \label{sec:coverage_probability}
This section analyzes the coverage probability of CoMP transmission in VHetNets. Let $\mathcal{C}$ be the event that a typical user is in coverage. Recalling the total probability formula, the coverage probability can be computed as
\begin{equation*}
  P = \sum_{\chi\in Q}\mathbb{P}(\mathcal{C}|\mathcal{E}_{\chi})\mathbb{P}(\mathcal{E}_{\chi}),
\end{equation*}
where the operator $\mathbb{P}(x)$ denotes the probability of event $x$ and $\mathbb{P}(x|y)$ denotes the conditional probability of event $x$ given~$y$.

Before proceeding to the coverage analysis, we first investigate the conditional distance distribution of a typical aerial UE, given its association with either three serving ABSs or three serving TBSs. The following lemmas then characterize the resulting distances to the three serving base stations in each case.
\begin{subequations}
\begin{lemma}\label{lem:condition_JPDF_distance}
  When an aerial UE is associated with either three ABSs or three TBSs (i.e., event $\mathcal{E}_{\ABS}$ or $\mathcal{E}_{\TBS}$ occurs), the joint PDF of the distances between the UE and the three ABSs or three TBSs is given, respectively, by
  \begin{align}
    f_{\boldsymbol{R}\mid\mathcal{E}_{\ABS}}^{\ABS}(\boldsymbol{r}) &= \frac{1}{\mathcal{A}_{\ABS}}\bar{F}_{\boldsymbol{R}}^{\TBS}(\boldsymbol{r})
    f_{\boldsymbol{R}}^{\ABS}(\boldsymbol{r}), \\
    f_{\boldsymbol{R}\mid\mathcal{E}_{\TBS}}^{\TBS}(\boldsymbol{r}) &= \frac{1}{\mathcal{A}_{\TBS}}\bar{F}_{\boldsymbol{R}}^{\ABS}(\boldsymbol{r})
    f_{\boldsymbol{R}}^{\TBS}(\boldsymbol{r}),
    \end{align}
  where
  \begin{align}
\bar{F}_{\boldsymbol{R}}^{\ABS}(\boldsymbol{r})  &= 1 - \int_{H-h}^{r_1}\int_{x}^{r_2} \int_{y}^{r_3} \hspace{-0.5em} f_{\boldsymbol{R}}^{\ABS}(x,y,z) \, \mathrm{d}z \, \mathrm{d}y \, \mathrm{d}x, \label{eq:R_ABS_CCDF} \\
\bar{F}_{\boldsymbol{R}}^{\TBS}(\boldsymbol{r})  &= 1 - \int_{h}^{r_1}\int_{x}^{r_2} \int_{y}^{r_3} \hspace{-0.5em} f_{\boldsymbol{R}}^{\TBS}(x,y,z) \, \mathrm{d}z \, \mathrm{d}y \, \mathrm{d}x.  \label{eq:R_TBS_CCDF}
  \end{align}
\end{lemma}
\end{subequations}

\begin{IEEEproof}
  Let $\boldsymbol{r} = (r_1, r_2, r_3)$ be an ordered distance vector with $H - h \leq r_1 < r_2 < r_3 \leq r_{\max}$. Define
$\boldsymbol{R}^{\ABS} = (R_1^{\ABS}, R_2^{\ABS}, R_3^{\ABS})$ and
$\boldsymbol{R}^{\TBS} = (R_1^{\TBS}, R_2^{\TBS}, R_3^{\TBS})$.
The conditional joint CDF of $R_1^{\ABS}$, $R_2^{\ABS}$, and $R_3^{\ABS}$ is given by
\begin{align*}
  F_{\boldsymbol{R}\mid\mathcal{E}_{\ABS}}^{\ABS}(\boldsymbol{r})&= \frac{1}{\mathcal{A}_{\ABS}} \mathbb{P}\left(\boldsymbol{R}^{\ABS} < \boldsymbol{r};\, \boldsymbol{R}^{\TBS} > \boldsymbol{r}\right) \\
  &= \frac{1}{\mathcal{A}_{\ABS}} F_{\boldsymbol{R}}^{\ABS}(\boldsymbol{r}) \bar{F}_{\boldsymbol{R}}^{\TBS}(\boldsymbol{r}) \\
  &= \frac{1}{\mathcal{A}_{\ABS}} \int\limits_{\substack{H - h \leq \boldsymbol{x} \leq \boldsymbol{r}}} \hspace{-1.5em}
     \bar{F}_{\boldsymbol{R}}^{\TBS}(\boldsymbol{r})
     f_{\boldsymbol{R}}^{\ABS}(\boldsymbol{x})\, \mathrm{d}\boldsymbol{x},
\end{align*}
where $\mathcal{E}_{\ABS} = \{\boldsymbol{R}^{\TBS} > \boldsymbol{r}\}$. Here, $\bar{F}_{\boldsymbol{R}}^{\TBS}$ denotes the joint CCDF given in~\eqref{eq:R_TBS_CCDF}. The result follows by applying the multivariate Leibniz rule for differentiation under the integral sign~\cite{rudin1976pma}. A similar derivation yields the conditional joint PDF~$F_{\boldsymbol{R}\mid\mathcal{E}_{\TBS}}^{\TBS}$.
\end{IEEEproof}

The Laplace transform of $\sqrt{I}$, serving as a key intermediate result for the coverage probability, can be derived as
  \begin{align}\label{eq:sqrt_I}
    \mathcal{L}_{\sqrt{I}\mid\mathcal{E}_\chi}(s) &= \mathbb{E}\left[\exp\left(-s\sqrt{I}\right)\right] \nonumber \\
    &= \int_0^\infty \frac{s}{2\sqrt{\pi} u^{3/2}} e^{-s^2/(4u)} \mathcal{L}_{I\mid\mathcal{E}_\chi}(u) \, \mathrm{d}u,
  \end{align}
 where \cite[Eq. (3.471.9)]{gradshteyn2007table} is applied to derive \eqref{eq:sqrt_I}, and $\mathcal{L}_{I\mid\mathcal{E}_\chi}(u)$ can be explicitly computed as \eqref{eq:expected_I}, shown at the bottom of the page,
 \begin{figure*}[!b]
 \hrulefill
\begin{align}
  \mathcal{L}_{I\mid\mathcal{E}_\chi}(u)
  &= \exp\Bigg(-2\pi\lambda_{\TBS}\int_{l_1(r_3)}^{\infty}\left[1-\left(\frac{m_{\TBS,\LoS}}{m_{\TBS,\LoS}+u\left(z^2+h^2\right)^{-\alpha_{\TBS,\LoS}}} \right)^{m_{\TBS,\LoS}} \right]zP_\LoS \, \mathrm{d}z \nonumber \\
  &\quad {}-2\pi\lambda_{\TBS}\int_{l_1(r_3)}^{\infty}\left[1-\left(\frac{m_{\TBS,\NLoS}}{m_{\TBS,\NLoS}+u\left(z^2+h^2\right)^{-\alpha_{\TBS,\NLoS}}} \right)^{m_{\TBS,\NLoS}} \right]zP_\NLoS \, \mathrm{d}z \nonumber \\
  &\quad {}-\left[\frac{2}{r_{C}^2}\int_{l_2(r_3)}^{r_{C}}\left(\frac{m_\ABS}{m_\ABS+u\left(z^2+(H-h)^2\right)^{-\alpha_\ABS}} \right)^{m_\ABS} z \, \mathrm{d}z \right]^{N-k} \Bigg).
  \label{eq:expected_I}
\end{align}
\end{figure*}
in which the parameters are defined as:
\begin{align*}
l_1(r_3) &=
\left\{\begin{array}{rl}
\sqrt{r_3^2 - h^2}, & \text{if } \chi = \TBS; \\
0,                  & \text{if } \chi = \ABS,
\end{array} \right. \\[1ex]
l_2(r_3) &=
\left\{\begin{array}{rl}
0,                             & \text{if } \chi = \TBS; \\
\sqrt{r_3^2 - (H - h)^2},      & \text{if } \chi = \ABS,
\end{array} \right. \\[1ex]
k &=
\left\{\begin{array}{rl}
0, & \text{if } \chi = \TBS; \\
3, & \text{if } \chi = \ABS.
\end{array} \right.
\end{align*}

We now present the main result for the overall coverage probability:
\begin{theorem}[Network Coverage Probability]\label{thm:coverage_probability}
The network coverage probability $P$ of a randomly located aerial UE in a VHetNet is given by the weighted sum of the conditional coverage probabilities under ABS and TBS association:
  \begin{equation}
    P = P_{\ABS}\mathcal{A}_{\ABS}+ P_\TBS\mathcal{A}_{\TBS},
  \end{equation}
 where the association probabilities $\mathcal{A}_{\ABS}$ and $\mathcal{A}_{\TBS}$ are given by \eqref{eq:probability_A_ABS} and \eqref{eq:probability_A_TBS}, respectively. The conditional coverage probabilities under the ABS-association event $\mathcal{E}_{\ABS}$ and the TBS-association event $\mathcal{E}_{\TBS}$ are given by
  \begin{align*}
    P_{\ABS}&=\hspace{-2em}\int\limits_{\substack{H - h \leq r_1 \leq r_2 \leq r_3 \leq r_{\max}}} \hspace{-2em}
    \sum_{k=0}^{\tilde{\nu}_{\ABS}-1} \frac{(-\sqrt{\gamma_{\ABS}})^k}{k!\theta_{\ABS}^k}\frac{\partial^{k}\mathcal{L}_{\sqrt{I}\mid\mathcal{E}_\ABS} (s)}{\partial s^{k}}\bigg|_{s=\frac{\sqrt{\gamma_{\ABS}}}{\theta_{\ABS}}} \\
    &\hspace{6em} \times f_{\boldsymbol{R}\mid\mathcal{E}_{\ABS}}^{\ABS}(\boldsymbol{r})\mathrm{d}\boldsymbol{r}, \\
    P_{\TBS}&=\hspace{-1em}\int\limits_{h\leq r_1\leq r_2 \leq r_3\leq \infty}\sum_{\boldsymbol{\zeta}\in \{\LoS,\NLoS\}^3}\left[\prod_{i=1}^{3}P_{\zeta_i}(r_i)\right] \\
    &\quad\sum_{k=0}^{\tilde{\nu}_{\TBS,\boldsymbol{\zeta}}-1} \frac{(-\sqrt{\gamma_{\TBS}})^k}{k!\theta_{\TBS,\boldsymbol{\zeta}}^k}\frac{\partial^{k}\mathcal{L}_{\sqrt{I}\mid\mathcal{E}_\TBS} (s)}{\partial s^{k}}\bigg|_{s=\frac{\sqrt{\gamma_{\TBS}}}{\theta_{\TBS,\boldsymbol{\zeta}}}} \hspace{-0.6em}f_{\boldsymbol{R}\mid\mathcal{E}_{\TBS}}^{\TBS}(\boldsymbol{r})\mathrm{d}\boldsymbol{r},
  \end{align*}
  where $\tilde{\nu}_{\ABS} = \mathrm{round}(\nu_\ABS)$ and $\tilde{\nu}_{\TBS,\boldsymbol{\zeta}} = \mathrm{round}(\nu_{\TBS,\boldsymbol{\zeta}})$, and the Laplace transform $\mathcal{L}_{\sqrt{I}\mid\mathcal{E}_\chi}(s)$ is given by \eqref{eq:sqrt_I}.
\end{theorem}

\begin{IEEEproof}
  See Appendix \ref{sec:proof_coverage_probability}.
\end{IEEEproof}

\begin{remark}[Altitude-Dependent Coverage Performance]
\label{remark:coverage_regimes}
While the association preference varies with user altitude, as established in Proposition~\ref{proposition:two_regimes}, the overall coverage probability also exhibits altitude-dependent regimes, although its dynamics are governed by distinct interference and fading characteristics.
\begin{itemize}
	\item \textbf{Low-altitude regime ($h < h_{\text{\rm th}}$)}: Even under frequent LoS blockage, the reduced path loss resulting from the shorter TBS link distance remains a key enabler of coverage. Additionally, the supplementary coverage provided by the ABS in lower-altitude regions mitigates the residual blockage effects. As a result, the overall network exhibits robust and satisfactory coverage performance.
	\item \textbf{High-altitude regime ($h > h_{\text{\rm th}}$)}: At higher altitudes, the network coverage primarily benefits from ABS connectivity. As LoS conditions become dominant and the path length to the ABSs shortens, the resulting favorable link budgets outweigh the impact of elevated co-channel interference, thereby enhancing the overall coverage performance.
\end{itemize}

This nuanced behavior highlights the importance of jointly optimizing user association strategies and physical-layer link performance in the design of altitude-aware VHetNet deployments, a promising research topic for the future.

\end{remark}

\section{ABS Deployment Optimization}
\label{sec:deployment_optimization}
The preceding analysis assumed randomly distributed base stations—a strategy well-suited for scenarios requiring rapid, uncoordinated deployment, such as disaster recovery or emergency coverage. While analytically tractable within the stochastic geometry framework, random deployment primarily serves as a performance baseline, providing theoretical insights into the average coverage probability and the spatial behavior of VHetNets.

In contrast, this section focuses on \emph{intentional ABS deployment}, tailored to address known coverage deficiencies in a given area. To this end, we propose a coverage-aware optimization framework that leverages a weighted $K$-means clustering algorithm in conjunction with CoMP transmission to guide ABS placement for maximal coverage efficiency.

Consider a set of user or sampling locations $\mathcal{X} = \{x_1, \cdots, x_N\} \subset \mathbb{R}^2$\, within a target region, along with a corresponding set of weights $\mathcal{W} = \{w_1, \cdots, w_N\}$ that quantify the severity of coverage shortfall at each location. The objective is to position $K$ ABSs at $\{\mu_k\}_{k=1}^K$ so as to maximize the weighted average success probability:
\begin{equation}
\label{eq:obj_fading}
  \max_{\{C_k\},\{\mu_k\}}
  \sum_{k=1}^{K} \sum_{x_i \in C_k}
  w_i \left(1 + \frac{1}{m} \lVert x_i - \mu_k \rVert^{\alpha} \right)^{-m},
\end{equation}
where $\alpha$ is the path loss exponent, $m$ is the Nakagami fading parameter, and $C_k$ is the cluster of points associated with the $k^{\text{th}}$ ABS, and $w_i = \max(0, \gamma_{\TBS} - \gamma_i)$ reflects the SIR gap at location $x_i$ relative to a threshold $\gamma_{\TBS}$. Locations experiencing poor terrestrial coverage (i.e., $\gamma_i < \gamma_{\TBS}$) receive higher weights, prioritizing them in the optimization.

The objective in \eqref{eq:obj_fading} generalizes several canonical formulations:
\begin{itemize}
  \item Rayleigh fading ($m=1$):
  \begin{equation}
  \max_{\{C_k\},\{\mu_k\}}
  \sum_{k=1}^K \sum_{x_i\in C_k}
  \frac{w_i}{1+ \lVert x_i-\mu_k\rVert^\alpha}.
  \end{equation}

  \item Deterministic path loss ($m\to\infty$):
  \begin{equation}
  \max_{\{C_k\},\{\mu_k\}}
  \sum_{k=1}^K \sum_{x_i\in C_k}
  w_i \, e^{-\lVert x_i-\mu_k\rVert^\alpha}.
  \end{equation}

  \item Free-space propagation ($\alpha=2,\,m\to\infty$):
  \begin{equation}
  \max_{\{C_k\},\{\mu_k\}}
  \sum_{k=1}^K \sum_{x_i\in C_k}
  w_i \, e^{-\lVert x_i-\mu_k\rVert^2},
  \end{equation}
  which behaves similarly to the classical weighted $K$-means objective
  \begin{equation}
  \min_{\{C_k\},\{\mu_k\}}
  \sum_{k=1}^K \sum_{x_i\in C_k}
  w_i \, \lVert x_i-\mu_k\rVert^2,
  \end{equation}
  since $\exp(- d^2)\approx 1- d^2$ for small $d^2$.
\end{itemize}
Thus, the objective function given by \eqref{eq:obj_fading} can be viewed as a fading- and path-loss-aware extension of the traditional $K$-means strategy, grounded in physical-layer considerations.

%In implementation, ABS centers $\{\mu_k\}$ are initialized via \emph{k-means++} on the hole set $\mathcal{H}=\{x_i:\gamma_i<\gamma_{\TBS}\}$ while passing the hole severity as sample weights (\texttt{sample\_weight} $= \gamma_{\TBS}-\gamma_i$). This spreads seeds across uncovered regions and biases toward deeper holes.

The deployment procedure is formalized in Algorithm~\ref{alg:fading_kmeans}. The algorithm alternates between: \emph{i)} assigning each $x_i$ to the ABS that maximizes its fading-aware kernel in \eqref{eq:obj_fading}, and \emph{ii)} updating $\mu_k$ as the weighted centroid of cluster $C_k$, where weights incorporate both $w_i$ and the fading-aware kernel. This process monotonically increases the objective and converges in a finite number of iterations.

\begin{algorithm}[!t]
\caption{\small Path-Loss \& Fading-Aware Clustering for UAV Deployment}
\label{alg:fading_kmeans}
\small
\begin{algorithmic}[1]
\Require
    $\mathcal{X}$: user/hole coordinates, \\
    $\mathcal{W}$: weights, \\
    $K$: number of UAVs, \\
    $\alpha$: path loss parameter, \\
    $m$: Nakagami-m fading parameter, \\
    $\epsilon$: convergence threshold, \\
    $T_{\max}$: max iterations
\Ensure
    Optimized UAV positions $\{\mu_k^*\}$

\State Initialize $\{\mu_k^{(0)}\}$

\For{$t = 0$ to $T_{\max}$}
    \For{each $x_i \in \mathcal{X}$}
        \State Assign $x_i$ to cluster
        \[
        k = \arg\max_j \; w_i \Big(1+\tfrac{1}{m} \lVert x_i-\mu_j^{(t)} \rVert^\alpha \Big)^{-m}
        \]
    \EndFor
    \For{$k = 1$ to $K$}
        \State Update cluster center:
        \[
        \mu_k^{(t+1)} =
        \frac{\sum_{x_i \in C_k}
        w_i \Big(1+\tfrac{1}{m} \lVert x_i - \mu_k^{(t)} \rVert^\alpha \Big)^{-m} x_i}
        {\sum_{x_i \in C_k}
        w_i \Big(1+\tfrac{1}{m}  \lVert x_i - \mu_k^{(t)} \rVert^\alpha \Big)^{-m}}
        \]
    \EndFor
    \If{All $\|\mu_k^{(t+1)}-\mu_k^{(t)}\| < \epsilon$}
        \State \textbf{break}
    \EndIf
\EndFor
\State \Return $\{\mu_k^*\}=\{\mu_k^{(t+1)}\}$
\end{algorithmic}
\end{algorithm}

This optimization framework ensures UAV-based ABSs are deployed exactly where they maximize users’ success probability, yielding substantial performance improvements over geometric clustering. It retains the efficiency of $K$-means while aligning the optimization objective with wireless coverage performance, enabling real-time adaptation in dynamic environments such as post-disaster recovery, rural connectivity, and hotspot offloading~\cite{8660516, 8675384, 7470933}.

\section{Numerical Results and Discussions} \label{sec:numerical_results}
This section presents Monte Carlo simulation results to validate the analytical expressions and examine the impact of key system parameters on the CoMP performance in VHetNets. Specifically, we assess how network coverage responds to variations in ABS/TBS densities, aerial user altitude, and spatial distributions of users.

By 3GPP guidelines \cite{3gpp_tr_36_777}, the maximum operational altitude for aerial users is set to $300$ \si{m}. For simplicity, a unified SIR threshold is assumed across all transmission scenarios, i.e., $\gamma_{\ABS} = \gamma_{\TBS} = \gamma$. Additionally, the path loss exponent is taken to be the same for A2A and LoS G2A links, i.e., $\alpha_{\ABS} = \alpha_{\TBS, \LoS} = \alpha$. The key simulation parameters are summarized in Table~\ref{tab2} for reference.

\begin{table}
\begin{center}
\renewcommand\arraystretch{1.25}
\caption{Simulation Parameter Setting}
\label{tab2}
\begin{tabular}{!{\vrule width1.2pt} c !{\vrule width1.2pt} c !{\vrule width1.2pt}}
\Xhline{1.2pt}
\textbf{Parameter} & \textbf{Value} \\
\Xhline{1.2pt}
$r_C$ & 1000 \si{m} \\
\hline
$h_{\TBS}$ & 30 \si{m} \\
\hline
$h$ & 120 \si{m} \\
\hline
$H$ & 320 \si{m} \\
\hline
$N$ & $20$ \\
\hline
$\lambda_{\TBS}$ & $20~\text{km}^{-2}$ \\
\hline
$\gamma$ & 0 \si{dB}\\
\hline
$(\alpha_{\ABS},\alpha_{\TBS,\LoS},\alpha_{\TBS,\NLoS})$ & $(2,2,2.7)$ \\
\hline
$(m_{\ABS},m_{\TBS,\LoS},m_{\TBS,\NLoS})$ & $(2,2,1)$ \\
\hline
Suburban $(a,b,c)$ & $(1,6.581,1)$ \\
\hline
Highrise urban $(a,b,c)$ & $(1.124,0.049,1.024)$ \\
\Xhline{1.2pt}
\end{tabular}
\end{center}
\end{table}

\begin{figure}[!t]
\vspace{-15pt}
  \centering
  \captionsetup[subfigure]{margin=5pt}
  \subfloat{%
    \includegraphics[width=0.5\linewidth]{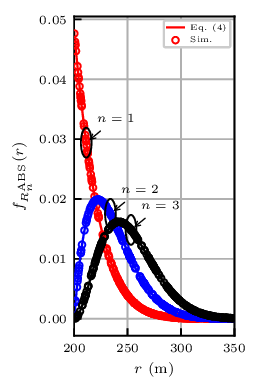}%
  }\hfill
  \subfloat{%
    \includegraphics[width=0.5\linewidth]{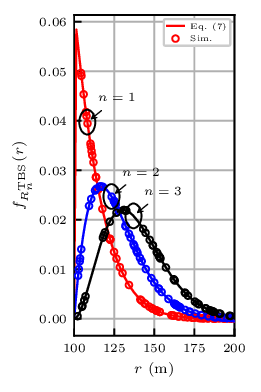}%
  }
  \vspace{-10pt}
  \caption{The accuracy of the PDFs $R_n^\ABS$ and $R_n^\TBS$, as defined in (\ref{eq:A2A_PDF_n_distance}) and (\ref{eq:G2AL_PDF_n_distance}), is validated by simulation results.}
  \label{fig:fi}
  \vspace{-10pt}
\end{figure}

\begin{figure}[!t]
  \centering
  \captionsetup[subfigure]{margin=5pt}
  \subfloat{%
    \includegraphics[width=0.5\linewidth]{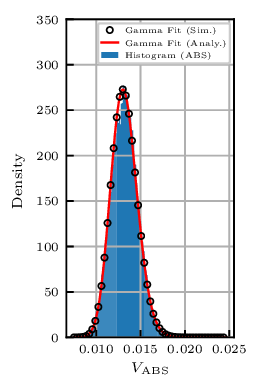}%
  }\hfill
  \subfloat{%
    \includegraphics[width=0.5\linewidth]{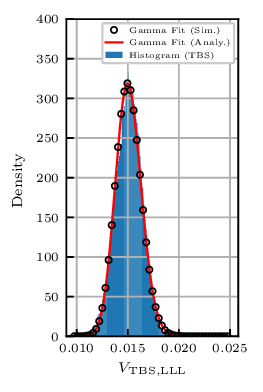}%
  }
  \vspace{-10pt}
  \caption{The accuracy of the PDFs $V_\ABS$ and $V_{\TBS,\LoS\LoS\LoS}$, defined in \eqref{eq:PDF_T_p}, is verified through comparison with simulation results.}
  \label{fig:f_Ti}
  \vspace{-10pt}
\end{figure}

\subsection{Distance and Received Signal Distributions}
Fig.~\ref{fig:fi} depicts the PDFs of the distances $R_n^{\ABS}$ and $R_n^{\TBS}$ between a typical user and the $n$-th closest ABS or TBS, respectively. The curves derived analytically align closely with the simulation results, thereby validating the correctness of the derived joint distance distributions.

Fig.~\ref{fig:f_Ti} shows the distributions of the received signal power $V_{\ABS}$ and the aggregate terrestrial signal $V_{\TBS, \LoS\LoS\LoS}$ under the fully-LoS configuration. Both empirical histograms and their respective Gamma distribution fittings are displayed. As illustrated, the Gamma approximation closely matches the empirical data, confirming its suitability for accurately modeling these random variables in analytical expressions.

\begin{figure}[!t]
  \centering
  \captionsetup[subfigure]{margin=5pt}
  \subfloat{%
    \includegraphics[width=0.5\linewidth]{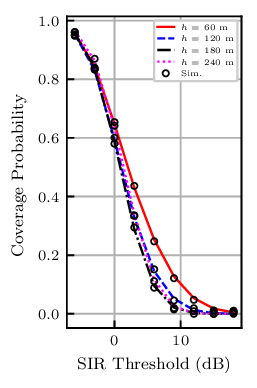}%
  }\hfill
  \subfloat{%
    \includegraphics[width=0.5\linewidth]{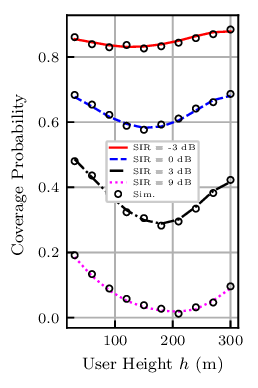}%
  }
  \vspace{-10pt}
  \caption{Coverage probability versus SIR threshold $\gamma$ and user altitude $h$, with ABSs fixed at $H=320$ \si{m}.}
  \label{fig:coverage_t_h}
  \vspace{-10pt}
\end{figure}

\subsection{Coverage Probability}
Fig.~\ref{fig:coverage_t_h} illustrates the coverage probability as a function of the SIR threshold $\gamma$ and the aerial user altitude $h$. In the left subplot, for all altitudes considered ($60-240$ \si{m}), the coverage probability decreases monotonically with increasing $\gamma$, which is consistent with standard SIR behavior. Notably, for $\gamma < 3$ \si{dB}, the curves across different altitudes nearly coincide, indicating minimal sensitivity to user height. However, when $\gamma > 3$ \si{dB}, the trajectory corresponding to $h = 60$ \si{m} achieves the highest coverage probability, while curves for higher altitudes remain similar but lower. This implies that, under mild SIR requirements, altitude has a negligible impact on coverage. In contrast, lower altitudes provide a distinct advantage under more stringent conditions.

The right subplot of Fig.~\ref{fig:coverage_t_h} examines coverage as a function of user altitude $h$ for several fixed SIR thresholds. The resulting curves are generally convex, suggesting that users positioned at either low or high altitudes experience better coverage than those at intermediate heights. This behavior aligns with the association policy discussed in Section~\ref{sec:association_policy}, where users tend to associate exclusively with either three ABSs or three TBSs, with mixed associations being rare. Interestingly, the altitudes corresponding to minimum and maximum coverage vary with $\gamma$. For instance, at $\gamma = -3$~\si{dB}, peak coverage occurs at $300$ \si{m} and the minimum around $120$~\si{m}; whereas for $\gamma = 9$ \si{dB}, the optimal altitude drops to $30$ \si{m}, with the minimum near $210$~\si{m}. This observation highlights the dynamic nature of coverage performance as a function of altitude and SIR threshold.

\begin{figure}[!t]
  \centering
  \captionsetup[subfigure]{margin=5pt}
  \subfloat{%
    \includegraphics[width=0.5\linewidth]{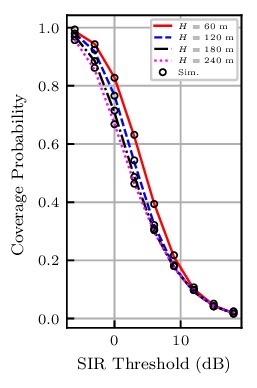}%
  }\hfill
  \subfloat{%
    \includegraphics[width=0.5\linewidth]{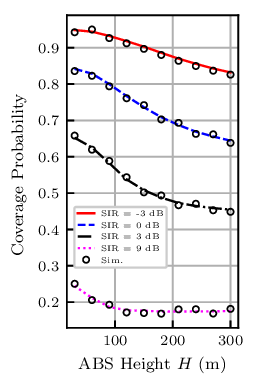}%
  }
  \vspace{-10pt}
  \caption{Coverage probability versus SIR threshold $\gamma$ and ABS altitude $H$, with the user fixed at $h=30$ \si{m}.}
  \label{fig:coverage_t_ABS}
  \vspace{-10pt}
\end{figure}

Fig.~\ref{fig:coverage_t_ABS} presents the impact of ABS altitude on coverage probability, assuming a fixed aerial user height of $h = 30$~\si{m}. As shown in the left subplot, increasing the ABS height leads to a gradual decline in coverage probability. This trend becomes more subdued at higher ABS altitudes, where the rate of change diminishes. In the right subplot, coverage probability is plotted against ABS altitude for several fixed values of $\gamma$. When $\gamma$ is either very low or high, the curves remain relatively flat, indicating insensitivity to ABS height. However, for moderate thresholds, particularly around $\gamma = 3$ \si{dB}, the coverage declines sharply over a mid-range of altitudes. This suggests that the coverage performance is most sensitive to ABS deployment height under moderate SIR conditions.

\begin{figure}[!t]
\centering
\includegraphics[width=3.25in]{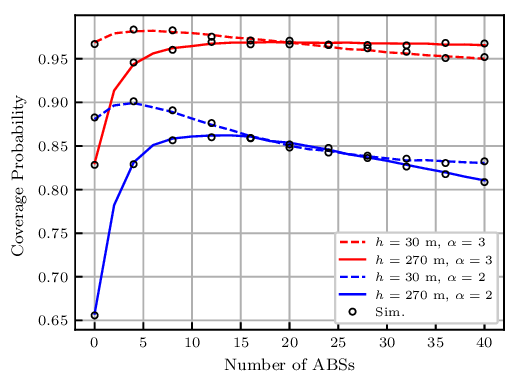}
\vspace{-10pt}
\caption{Coverage probability versus the number of ABSs under different user altitudes and path loss exponents.}
\label{fig:Coverage_lambda}
\vspace{-10pt}
\end{figure}

Fig.~\ref{fig:Coverage_lambda} illustrates the variation in coverage probability with the number of ABSs for different user altitudes and path loss exponents, where $\alpha_{\ABS} = \alpha_{\TBS,\LoS} = \alpha$. All curves exhibit a concave shape, with coverage probability increasing at low ABS numbers. This is because, at lower ABS counts, the increase in aerial interference power is relatively small, and the improvement in the desired aerial signal is more significant. Conversely, at higher ABS numbers, the coverage probability gradually decreases. This decline is due to the growing interference from additional ABSs, which outweighs the benefit of signal improvement, leading to a reduction in coverage probability.

These results suggest that, for a given path loss and user altitude, there exists an optimal number of UAVs for deployment. Beyond this optimal point, increasing the number of ABSs results in diminishing returns due to the increased interference, which reduces the overall coverage probability. Therefore, optimizing the number of ABSs is crucial for maximizing system performance, particularly in high-density or high-demand areas.

\subsection{Comparison with the Conventional Schemes}
\begin{figure}[!t]
\centering
\includegraphics[width=3.25in]{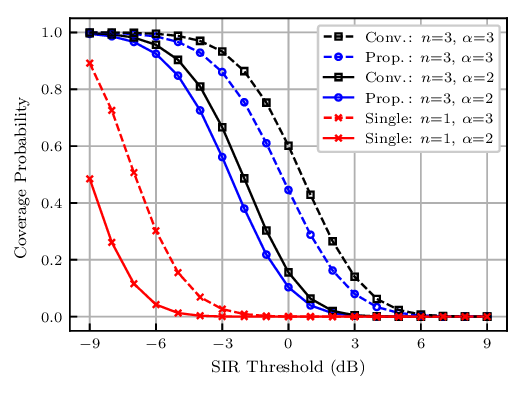}
\vspace{-10pt}
\caption{Coverage comparison among the single-link scheme in \cite{9250029}, the proposed CoMP scheme, and the strongest-three-BS rule under different path loss exponents.}\label{fig:Coverage_CoMP_VS_Single}
\vspace{-10pt}
\end{figure}

Fig.~\ref{fig:Coverage_CoMP_VS_Single} presents the coverage probability as a function of the SIR threshold $\gamma$ for two representative path loss exponents: $\alpha = 2$ and $\alpha = 3$. Three transmission schemes are compared:
\begin{enumerate}
    \item The proposed CoMP-based association strategy;
    \item A single-link baseline following the model in~\cite{9250029};
    \item A conventional heuristic that connects the user to the three strongest received-power base stations.
\end{enumerate}

Across the entire range of SIR thresholds, the proposed CoMP strategy consistently outperforms the single-link baseline and roughly matches the performance of the strongest-three heuristic. The improvement over single-link association is especially pronounced at moderate SIR thresholds. For instance, at $\gamma = -4$ \si{dB} with $\alpha = 3$, the coverage probability rises from approximately $0.1$ (single-link) to $0.9$ (CoMP). Similar gains are observed when $\alpha = 2$.

While the strongest-three heuristic achieves comparable coverage performance, it requires an exhaustive search over all potential base stations and a dynamic ranking of received power levels. In contrast, the proposed CoMP framework achieves equivalent or superior coverage with lower complexity, leveraging spatially optimized base station coordination and interference mitigation. This demonstrates its practical advantage, especially under dense deployments and harsh propagation environments.

\subsection{System-Level Performance Comparison}

\begin{figure}[t]
  \centering
  \subfloat[\scriptsize TBS Only.]{
    \includegraphics[width=0.46\linewidth]{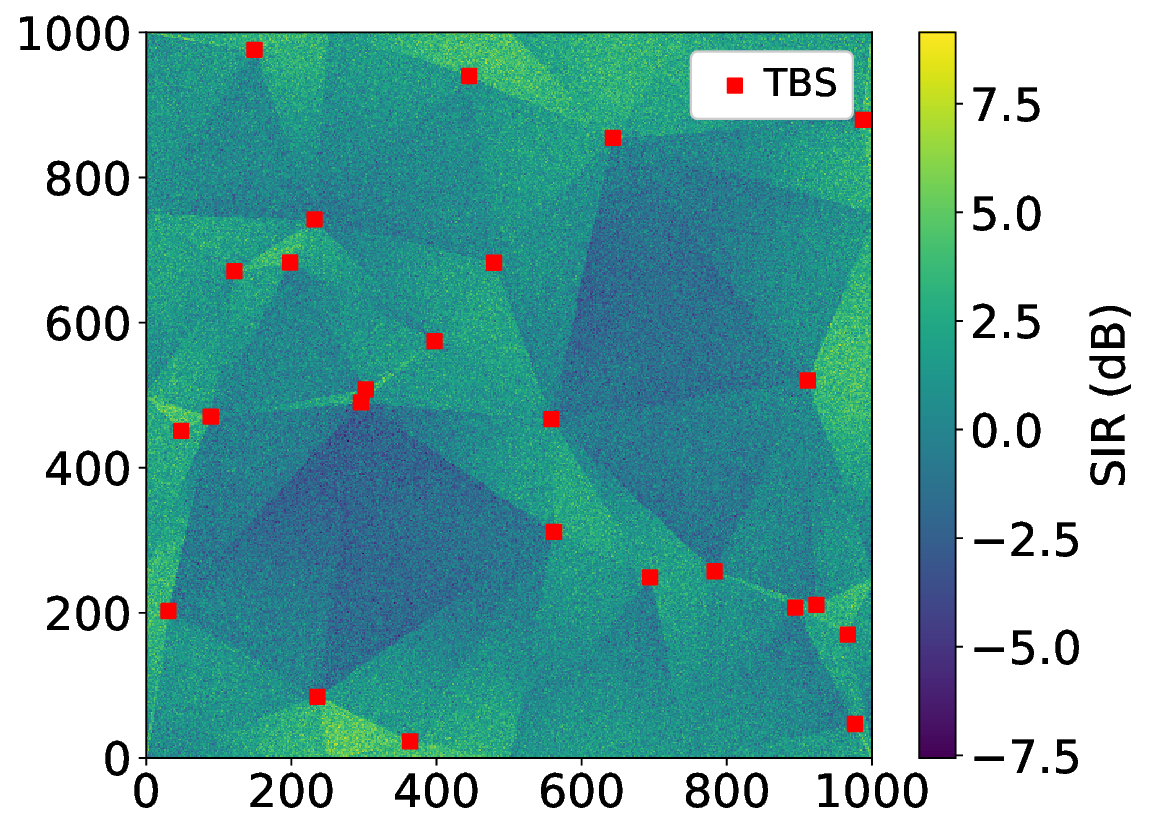}
    \label{subfig:SIR_heat-a}}
  \hspace{0.01\linewidth}
  \subfloat[\scriptsize TBSs with randomly deployed ABSs.]{
    \includegraphics[width=0.46\linewidth]{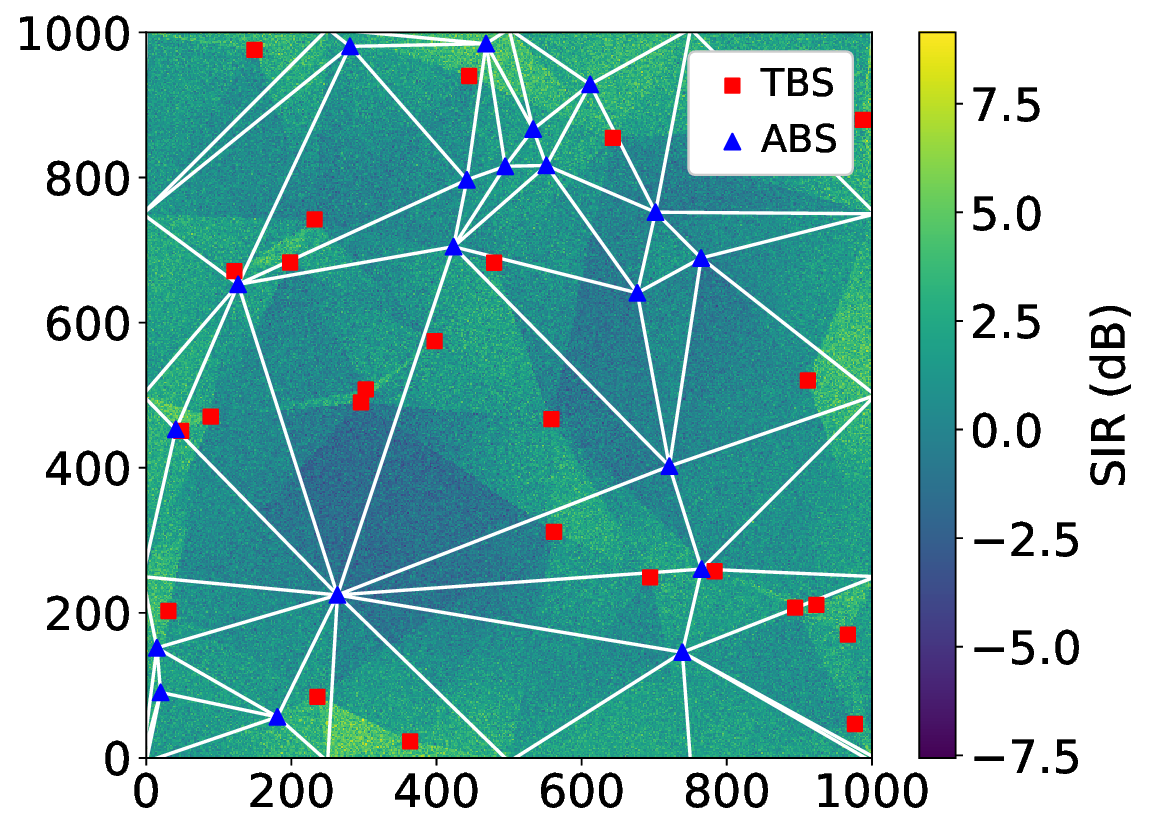}
    \label{subfig:SIR_heat-b}}
  \\
  \subfloat[\scriptsize TBSs with intentionally deployed ABSs by classical $K$-means algorithm.]{
    \includegraphics[width=0.46\linewidth]{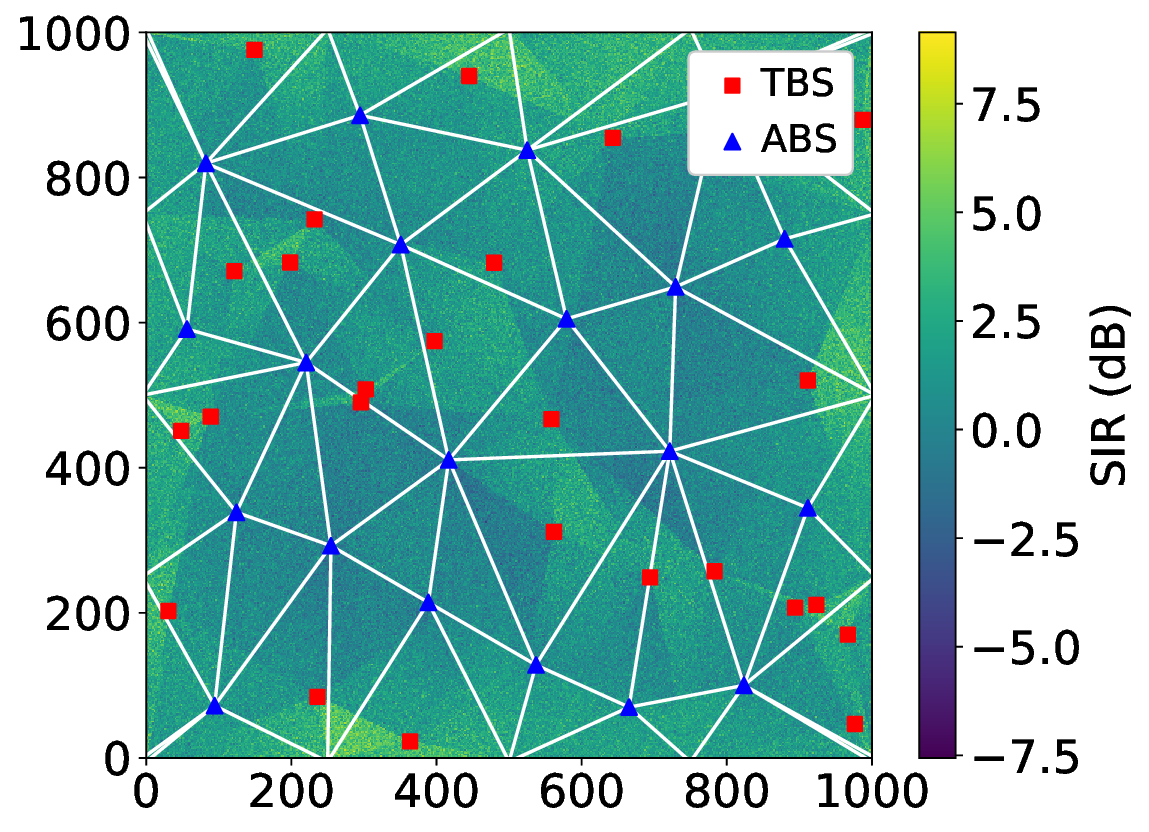}
    \label{subfig:SIR_heat-c}}
  \hspace{0.01\linewidth}
  \subfloat[\scriptsize TBSs with intentionally deployed ABSs by the proposed Algorithm~1.]{
    \includegraphics[width=0.46\linewidth]{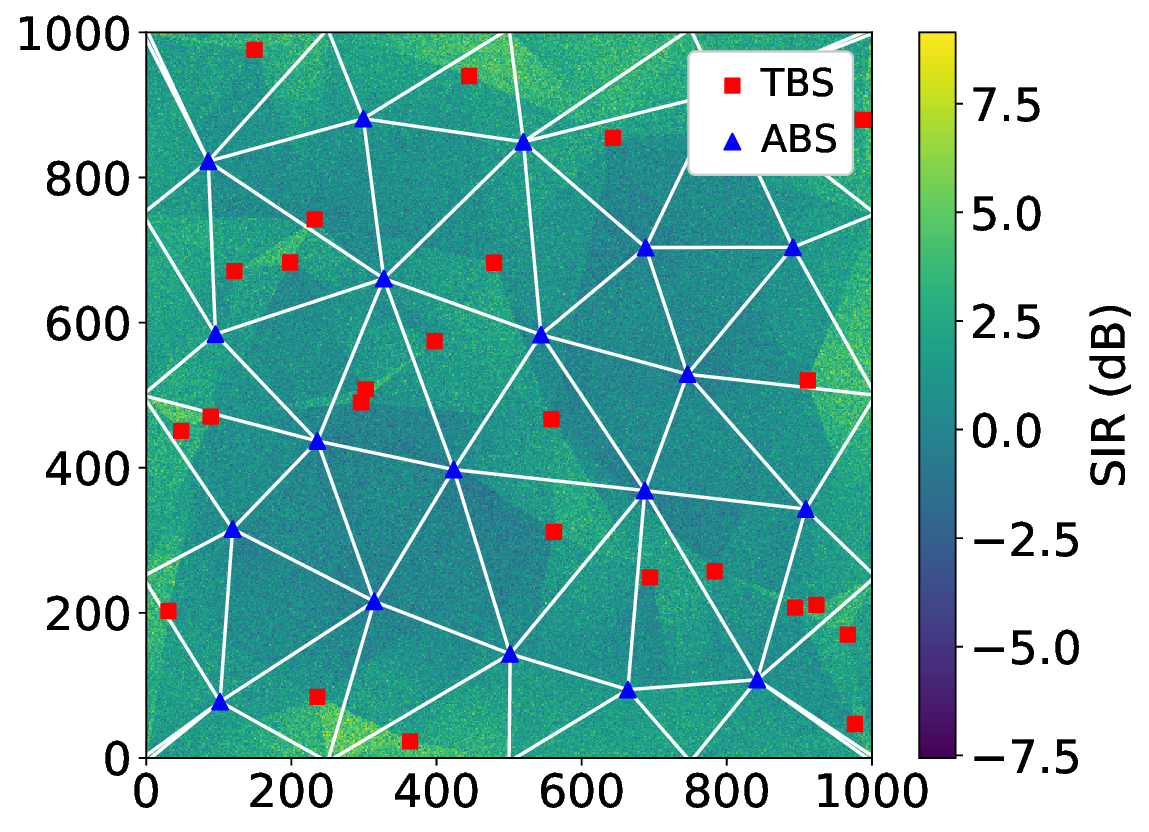}
    \label{subfig:SIR_heat-d}}
  \caption{Coverage heatmaps for four distinct deployment strategies.}
  \label{fig:SIR_heat}
\end{figure}

Fig.~\ref{fig:SIR_heat} compares four deployment strategies: {\it i)} TBS-only, {\it ii)} TBSs with randomly placed ABSs, {\it iii)} TBSs with ABSs positioned via the classical weighted $K$-means algorithm, and {\it iv)} TBSs with ABSs positioned using the proposed Algorithm~\ref{alg:fading_kmeans}. The heatmaps in Fig.~\ref{fig:SIR_heat} visually illustrate the coverage probabilities, with lighter colors indicating higher coverage and darker colors indicating lower coverage.

Progressing from Fig.~\ref{subfig:SIR_heat-a} to \ref{subfig:SIR_heat-d}, a clear trend of color lightening is observed, indicating improved coverage. The TBS-only configuration in Fig.~\ref{subfig:SIR_heat-a} exhibits relatively darker colors, corresponding to a coverage probability of $61.99\%$. With randomly deployed ABSs (Fig.~\ref{subfig:SIR_heat-b}), coverage improves to $72.93\%$, as reflected by lighter colors. The classical weighted $K$-means (Fig.~\ref{subfig:SIR_heat-c}) and our proposed Algorithm~\ref{alg:fading_kmeans} (Fig.~\ref{subfig:SIR_heat-d}) yield the lightest and most spatially uniform heatmaps, achieving the highest coverage probabilities of $79.85\%$ and $81.42\%$, respectively.

This improvement stems from geometry-aware optimization rather than random diversity. By weighting local SIR deficiencies, Algorithm~\ref{alg:fading_kmeans} reshapes the Voronoi partitions to approximate an acute Delaunay triangulation, which tends to minimize circumcircle radii \cite{6515119}. Indeed, Figs.~\ref{subfig:SIR_heat-c} and \ref{subfig:SIR_heat-d} contain far fewer obtuse triangles than Fig.~\ref{subfig:SIR_heat-b}, increasing the likelihood that user locations fall within strong cooperative coverage zones. Consequently, the optimized ABS deployment more effectively addresses coverage gaps, significantly enhancing the spatial efficiency and reliability of the network.

\section{Conclusion}\label{sec:conclusions}
This paper presented a comprehensive analysis and optimization framework for downlink coverage in CoMP-enabled VHetNets. A novel 3D network model based on Poisson–Delaunay triangulation was introduced to facilitate cooperative transmission among ABSs and TBSs. Using tools from stochastic geometry, we derived closed-form expressions for association probabilities, distance distributions, and coverage probabilities. Monte Carlo simulation experiments, conducted by 3GPP guidelines, validated the theoretical models and identified key performance trends related to user altitude, ABS height, base station density, and channel conditions. In particular, the proposed CoMP strategy markedly improved coverage over conventional single-link and heuristic methods, especially under challenging propagation environments. Furthermore, a coverage-aware weighted $K$-means deployment algorithm was shown to significantly outperform random UAV placement by targeting coverage-deficient regions. These findings underscore the benefits of cooperative transmission and geometry-aware UAV deployment in enhancing the reliability of low-altitude wireless access. Future research will integrate reinforcement learning with stochastic geometry to enable adaptive and real-time UAV positioning, supporting resilient, scalable, and self-organizing VHetNet infrastructures.

\appendices
\section{Proof of Lemma \ref{lem:A2A_PDF_n_distance}}\label{sec:proof_A2A_PDF_n_distance}
The distance from a typical aerial user located at the origin to its $n^\text{th}$ closest ABS is denoted by $ D_n $. According to \cite{5299075}, the CDF and PDF of $ D_n $ are respectively given by
\[
F_{D_n}(r) = \frac{r^2}{r_C^2}, \quad f_{D_n}(r) = \frac{2r}{r_C^2},
\]
where $ H - h \leq r \leq r_{\max} $.

The user associates with the $n^\text{th}$ closest base station. The CDF of $R_n^{\text{\rm ABS}}$ can be computed as
\begin{align*}
  F_{R_n}^{\ABS}(r)&= \mathbb{P}(R_n^{\ABS}<r) \\
   &= \mathbb{P}(\text{at least $n$ of the $R_n^{\ABS}$ are less than $r$}) \\
   &= \sum_{k=n}^{N}\binom{N}{k}\left(F_{D_n}(r)\right)^k(1-F_{D_n}(r))^{N-k}.
\end{align*}
By differentiating the above expression with respect to $r$, the PDF of the serving distance is obtained as
\begin{align} \label{eq:PDF_R_ABS1}
  \lefteqn{f_{R_n}^{\ABS}(r)
  = f_{D_n}\sum_{k=n}^{N}\binom{N}{k}} \nonumber \\
  &\quad \times\left[kF_{D_n}^{k-1}(1-F_{D_n})^{N-k}-(N-k)F_{D_n}^k(1-F_{D_n})^{N-k-1}\right].
\end{align}

\begin{figure}
  \centering
  \includegraphics[width=3.0in]{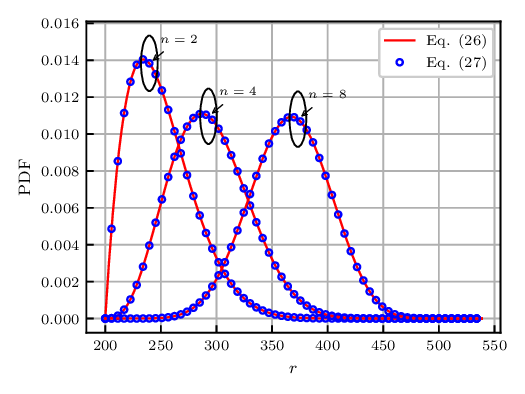}
  \vspace{-15pt}
  \caption{Two different representations of the PDF of the variable $R_n^\ABS$.}
  \label{fig:PDF_R_ABS1v2}
   \vspace{-10pt}
\end{figure}

Clearly, \eqref{eq:PDF_R_ABS1} is overly complex and not convenient for analytical or numerical computation. According to the theory of order statistics, a more concise closed-form expression can be obtained as follows \cite{david2004order}
\begin{align}\label{eq:PDF_R_ABS2}
f_{R_n}^{\ABS}(r) = \frac{N!}{(n-1)! (N-n)!} f_{D_n} \left[F_{D_n}\right]^{n-1} \left[1 - F_{D_n}\right]^{N-n}.
\end{align}
Substituting $F_{D_n}$ and $f_{D_n}$ into \eqref{eq:PDF_R_ABS2} yields \eqref{eq:A2A_PDF_n_distance} in Lemma \ref{lem:A2A_PDF_n_distance}.
To demonstrate the equivalence of \eqref{eq:PDF_R_ABS1} and \eqref{eq:PDF_R_ABS2}, a simulation was conducted with parameters set to $N=20$, $r_C=500$~\si{m}, $h=100$~\si{m}, and $H=300$~\si{m}. As shown in Fig. \ref{fig:PDF_R_ABS1v2}, the resulting curves from the two expressions perfectly coincide, thereby confirming their analytical equivalence.

According to the theory of order statistics \cite{david2004order}, the joint PDF of $R_{n_1}^{\ABS}, R_{n_2}^{\ABS}, \cdots, R_{n_k}^{\ABS}$ ($1\leq n_1<\cdots<n_k\leq N;1\leq k\leq n$) is for $r_1\leq\cdots\leq r_k$,
\begin{align*}
  &f_{R_{n_1},R_{n_2},\cdots,R_{n_k}}^{\ABS}(r_1, r_2,\cdots, r_k)= \\
&\frac{N!}{(n_1 - 1)! (n_2 - n_1 - 1)! \cdots (N - n_k)!} \prod_{i=1}^{k} f_{D_{n_i}}(r_i)  \\
&\times F_{D_{n_1}}^{n_1-1}(r_1)\left[ F_{D_{n_2}}(r_2) - F_{D_{n_1}}(r_1) \right]^{n_2 - n_1 - 1} \times \cdots \times \\
& \left[ F_{D_{n_k}}(r_k) - F_{D_{n_{k-1}}}(r_{k-1}) \right]^{n_k - n_{k-1} - 1}\left[1 - F_{D_{n_k}}(r_k) \right]^{N - n_k} \hspace{-0.5em}.
\end{align*}
For a sample of $N$ i.i.d.\ continuous random variables with PDF $f_{D_i}(x)$ and CDF $F_{D_i}(x)$, the joint PDF of the first $n$ order statistics $R_{1}^{\ABS} \leq R_{2}^{\ABS} \leq \cdots \leq R_{n}^{\ABS}$ is for $H-h\leq r_1\leq r_2\leq \cdots \leq r_n\leq r_{\max}$,
\begin{align*}
&f_{R_{1}, R_{2},\cdots , R_{n}}^{\ABS}(r_1, r_2,\cdots, r_n) = \\
&\frac{N!}{(N - n)!} \, f_{D_1}(r_1) f_{D_3}(r_2)\cdots f_{D_n}(r_n) \left[1 - F_{D_n}(r_n)\right]^{N - n}.
\end{align*}
Finally, substituting $f_{D_1},f_{D_2},\cdots,f_{D_n}$ and $F_{D_n}$ into the above expression yields \eqref{eq:A2A_JPDF_distance} in Lemma \ref{lem:A2A_PDF_n_distance}.

\section{Proof of Lemma \ref{lem:G2A_PDF_n_distance}}\label{sec:proof_G2A_PDF_n_distance}
The probability that less than $n$ nodes are closer than $r$ is:
\begin{align*}
  P_n&=\mathbb{P}(0, \cdots, n-1 \text{ nodes within } r) \\
  &= \sum_{k=0}^{n-1}\frac{(\lambda B_m(r))^k}{k!} e^{\lambda B_m(r)},
\end{align*}
where $B_m(r)$ is the area of the circle of radius $r$.

The CDF of $R_n^{\TBS}$ can be expressed as:
\begin{align} \label{eq:CDF_R_TBS}
  F_{R_n}^{\TBS}(r)&=1-\sum_{k = 0}^{n - 1}\frac{\left(\pi\lambda_{\TBS}(r^{2}-h^{2})\right)^k}{k!} \nonumber\\
  &\quad\times\exp\left(\pi\lambda_{\TBS}(r^{2}-h^{2})\right).
\end{align}
Taking differential of $F_{R_n}^{\TBS}(r)$ with respect to $r$ yields \eqref{eq:G2AL_PDF_n_distance}.

The PDF of distance $R_1^{\TBS}$ to the nearest TBS is
\begin{align*}
  f_{R_1}^{\TBS}(r_1)&=2\pi\lambda_{\TBS}r_1 \exp\left(-\pi\lambda_{\TBS}\left(r_1^{2}-h^{2}\right)\right).
\end{align*}
Given $R_1^{\TBS}=r_1$, the PDF of distance $R_2^{\TBS}$ to the second nearest base station is
\begin{align*}
  f_{R_2 \mid R_1}^{\TBS}(r_2\mid r_1)=2\pi\lambda_{\TBS}r_2 \exp\left(-\pi\lambda_{\TBS}\left(r_2^{2}-r_1^{2}\right)\right).
\end{align*}
By analogy, given $R_{n-1}^{\TBS}=r_{n-1}$, the PDF of distance $R_n^{\TBS}$ to the $n^\text{th}$ nearest base station is
\begin{align*}
  f_{R_n|R_{n-1}}^{\TBS}(r_n\mid r_{n-1})=2\pi\lambda_{\TBS}r_n\exp\left(-\pi\lambda_{\TBS}\left(r_n^{2}-r_{n-1}^{2}\right)\right).
\end{align*}
The joint PDF can be factorized using conditional probability as:
\begin{align*}
  &f_{R_1, R_2,\cdots, R_n}^{\TBS}(r_1,r_2,\cdots,r_n) \\
  &=f_{R_1}^{\TBS}(r_1)\times f_{R_2|R_1}^{\TBS}(r_2\mid r_1)\times\cdots\times f_{R_n|R_{n-1}}^{\TBS}(r_n\mid r_{n-1}) \\
  &=(2\pi\lambda_{\TBS})^nr_1r_2\cdots r_n \exp\left(-\pi\lambda_{\TBS}\left(r_n^{2}-h^{2}\right)\right).
\end{align*}

\section{Proof of Lemma \ref{lem:PDF_T}}\label{sec:proof_PDF_T}
The distribution of $U_{\chi,\boldsymbol{\zeta}} = \sum_{n\in \mathcal C_\chi}\left|{\rm H}^{(\chi,\zeta_n)}_{n}\right|(R_n)^{-\alpha_{\chi,\zeta_n}/2}$ admits a Gamma approximation (\ref{eq:PDF_T}) through the generalized central limit theorem \cite{papoulis1963fourier}, with shape and rate parameters defined explicitly as
\begin{equation*}
  \nu_{\chi,\boldsymbol{\zeta}} = \frac{\mathbb{E}^2[U_{\chi,\boldsymbol{\zeta}}]}{\text{Var}(U_{\chi,\boldsymbol{\zeta}})}, ~~ \theta_{\chi,\boldsymbol{\zeta}} = \frac{\text{Var}(U_{\chi,\boldsymbol{\zeta}})}{\mathbb{E}[U_{\chi,\boldsymbol{\zeta}}]},
\end{equation*}
where $\mathbb{E}[\cdot]$ and $\text{Var}(\cdot)$ denote the expectation and variance, respectively.

Since $\left|{\rm H}^{(\chi,\zeta_n)}_{n}\right|$ and $R_n$ are independence of each other, the expectation of $U_{\chi,\boldsymbol{\zeta}}$ can be written as
\begin{equation*}
  \mathbb{E}[U_{\chi,\boldsymbol{\zeta}}]= \sum_{n=1}^3\mathbb{E}\left[\left|{\rm H}^{(\chi,\zeta_n)}_{n}\right|\right]\mathbb{E}\left[(R_n)^{-\alpha/2}\right],
\end{equation*}
and the variance of $U_{\chi,\boldsymbol{\zeta}}$ is
\begin{align*}
  \text{Var}[U_{\chi,\boldsymbol{\zeta}}]&= \sum_{n=1}^3\mathbb{E}\left[\left|{\rm H}^{(\chi,\zeta_n)}_{n}\right|^2\right]\mathbb{E}[(R_n)^{-\alpha}] - \mathbb{E}^2[U_{\chi,\boldsymbol{\zeta}}] \\
  &\quad+\sum_{p\neq q}^3\mathbb{E}\left[\left|{\rm H}^{(\chi,\zeta_p)}_{p}\right|\right]\mathbb{E}\left[\left|{\rm H}^{(\chi,\zeta_q)}_{q}\right|\right]\mathbb{E}[(R_pR_q)^{-\alpha/2}],
\end{align*}
where $\mathbb{E}\left[\left|{\rm H}^{(\chi,\zeta_n)}_{n}\right|\right] = \frac{\Gamma(m_{\chi,\zeta_n} + \frac{1}{2})}{\Gamma(m_{\chi,\zeta_n})} \left(\frac{\Omega}{m_{\chi,\zeta_n}}\right)^{\frac{1}{2}}$ and $\mathbb{E}\left[\left|{\rm H}^{(\chi,\zeta_n)}_{n}\right|^2\right] = \Omega$.

\section{Proof of Theorem \ref{thm:coverage_probability}}\label{sec:proof_coverage_probability}
Suppose that a typical user is associated with three ABSs. The conditional coverage probability $P_{\ABS}$ is given by
  \begin{align*}
    P_{\ABS}&=\int_{\boldsymbol{r}>0} \mathbb{P}(\Gamma_{\ABS}>\gamma_{\ABS}|\boldsymbol{r}) \\
    &\quad\times f_{R_1, R_2, R_3|\mathcal{E}_{\ABS}}^{\ABS}(r_1,r_2,r_3) \, \mathrm{d}\boldsymbol{r}.
    \end{align*}
The coverage probability can be expressed as
    \begin{align*}
    \mathbb{P}\left(\Gamma_{\ABS} >\gamma_{\ABS}|\boldsymbol{r}\right)
    &= \mathbb{P}\left(\frac{S_{\ABS}}{I}>\gamma_{\ABS}|\boldsymbol{r}\right) \\
    &= \mathbb{E}_I\left[\mathbb{P}(S_{\ABS}>\gamma_{\ABS} I|\boldsymbol{r},I)\right].
  \end{align*}
The PDF, CDF, and CCDF of $ S_{\ABS} = U_{\ABS}^2 $ can be approximately expressed as
\begin{align*}
  f_{S_{\ABS}}(x) &\approx \frac{x^{(\nu_{\ABS} - 2)/2}}{2 \Gamma(\nu_{\ABS}) \theta_{\ABS}^{\nu_{\ABS}}}\exp\left(-\frac{\sqrt{x}}{\theta_{\ABS}}\right), \\
  F_{S_{\text{\rm ABS}}}(x) &\approx \frac{\gamma\left(\nu_{\text{\rm ABS}}, \frac{\sqrt{x}}{\theta_{\text{\rm ABS}}}\right)}{\Gamma(\nu_{\text{\rm ABS}})},  \\
  \bar{F}_{S_{\text{\rm ABS}}}(x) &\approx \frac{\Gamma\left(\nu_{\text{\rm ABS}}, \frac{\sqrt{x}}{\theta_{\text{\rm ABS}}}\right)}{\Gamma(\nu_{\text{\rm ABS}})},
\end{align*}
where $\gamma(\nu_{\text{\rm ABS}},z)=\int_{0}^{z}t^{\nu_{\text{\rm ABS}}-1}e^{-t} \, {\rm d}t$ and $\Gamma(\nu_{\text{\rm ABS}},z)=\int_{z}^{\infty}t^{\nu_{\text{\rm ABS}}-1}e^{-t} \, {\rm d}t$ are the lower and upper imcomplete Gamma functions, respectively. Next, we have
\begin{align*}
    &\mathbb{P}(\Gamma_{\text{\rm ABS}}>\gamma_{\text{\rm ABS}}|\boldsymbol{r}) \\
    &= \mathbb{E}_I \left[\frac{\Gamma(\nu_{\text{\rm ABS}},\frac{\sqrt{\gamma_{\text{\rm ABS}} I}}{\theta})}{\Gamma(\nu_{\text{\rm ABS}})}\right] \\
    &= \mathbb{E}_I \left[\exp(-\frac{\sqrt{\gamma_{\text{\rm ABS}} I}}{\theta})\sum_{k=0}^{\tilde{\nu}_{\ABS}-1}\frac{(\sqrt{\gamma_{\ABS} I})^k}{k!\theta^k}\right] \\
    &= \sum_{k=0}^{\tilde{\nu}_{\ABS}-1}\frac{(\sqrt{\gamma_{\ABS}})^k}{k!\theta^k}\mathbb{E}_I \left[\exp(-\frac{\sqrt{\gamma_{\ABS} I}}{\theta})(\sqrt{I})^k\right] \\
    &= \sum_{k=0}^{\tilde{\nu}_{\ABS}-1} \frac{(-\sqrt{\gamma_{\ABS}})^k}{k!\theta^k}\frac{\partial^{k}\mathcal{L}_{\sqrt{I}\mid\mathcal{E}_\ABS} (s)}{\partial s^{k}}\bigg|_{s=\frac{\sqrt{\gamma_{\ABS}}}{\theta}},
  \end{align*}
where $\tilde{\nu}_{\ABS} = \mathrm{round}(\nu_{\ABS})$ and $\mathcal{L}_{\sqrt{I}\mid\mathcal{E}_\ABS}(s)$ is given in~\eqref{eq:sqrt_I}. A similar derivation yields the conditional probability $P_{\TBS}$.

\bibliographystyle{IEEEtran.bst}
\bibliography{MyRef.bib}

\begin{IEEEbiography}
	[{\includegraphics[width=1in,height=1.25in, clip, keepaspectratio]{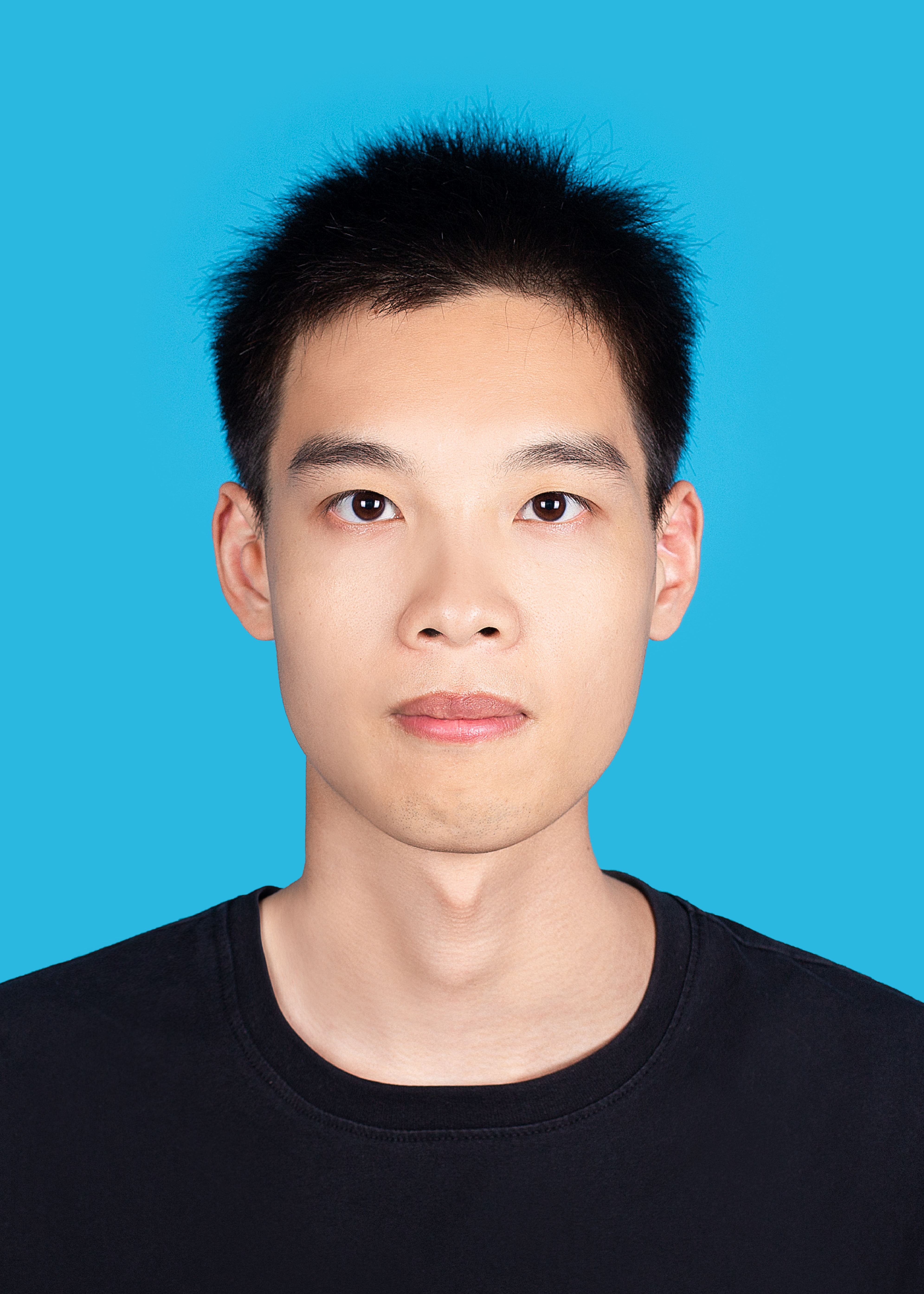}}]{Tian Shi} received the B.Sc. degree in information and computing science in 2019 and the M.Sc. Degree in mathematics in 2023, both from Guilin University of Electronic Technology, Guilin, China. He is currently pursuing a Ph.D. degree in information and communication engineering at Sun Yat-sen University, Guangzhou, China. His research interests include stochastic geometry and cooperative UAV communications.
\end{IEEEbiography}

\vfill

\begin{IEEEbiography}
	[{\includegraphics[width=1in, height=1.25in, clip, keepaspectratio]{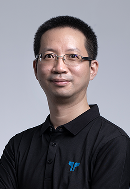}}]{Wenkun Wen} (Member, IEEE) received the Ph.D. degree in Telecommunications and Information Systems from Sun Yat-sen University, Guangzhou, China, in 2007. Since 2020, he has been with Techphant Technologies Co. Ltd., Guangzhou, China, as Chief Engineer.

From 2008 to 2009, he was with the Guangdong-Nortel R\&D center in Guangzhou, China, where he worked as a system engineer for 4G systems. From 2009 to 2012, he worked at the LTE R\&D center of New Postcom Equipment Co. Ltd., Guangzhou, China, where he served as the 4G standard team manager. From 2012 to 2018, he was with the 7th Institute of China Electronic Technology Corporation (CETC) as an expert in wireless communications. From 2018 to 2020, he served as Deputy Director of the 5G Innovation Center at CETC. His research interests include 5G/B5G mobile communications, machine-type communications, narrow-band wireless communications, and signal processing.
\end{IEEEbiography}	

\begin{IEEEbiography}
	[{\includegraphics[width=1in, height=1.25in, clip, keepaspectratio]{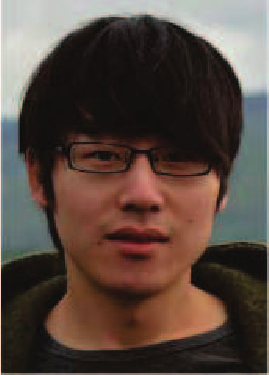}}]{Peiran Wu} (Member, IEEE) received the Ph.D. degree in electrical and computer engineering from The University of British Columbia (UBC), Vancouver, Canada, in 2015.
	
	From October 2015 to December 2016, he was a Post-Doctoral Fellow at UBC. In the Summer of 2014, he was a Visiting Scholar with the Institute for Digital Communications, Friedrich-Alexander-University Erlangen-Nuremberg (FAU), Erlangen, Germany. Since February 2017, he has been with Sun Yat-sen University, Guangzhou, China, where he is currently an Associate Professor. Since 2019, he has been an Adjunct Associate Professor with the Southern Marine Science and Engineering Guangdong Laboratory, Zhuhai, China. His research interests include mobile edge computing, wireless power transfer, and energy-efficient wireless communications. 
\end{IEEEbiography}

\begin{IEEEbiography}
	[{\includegraphics[width=1in, height=1.25in, clip, keepaspectratio]{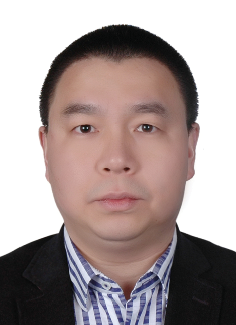}}]{Minghua Xia} (Senior Member, IEEE) received the Ph.D. degree in Telecommunications and Information Systems from Sun Yat-sen University, Guangzhou, China, in 2007.
	
	From 2007 to 2009, he was with the Electronics and Telecommunications Research Institute (ETRI) of South Korea, Beijing R\&D Center, Beijing, China, where he worked as a member and then as a senior member of the engineering staff. From 2010 to 2014, he was in sequence with The University of Hong Kong, Hong Kong, China; King Abdullah University of Science and Technology, Jeddah, Saudi Arabia; and the Institut National de la Recherche Scientifique (INRS), University of Quebec, Montreal, Canada, as a Postdoctoral Fellow. Since 2015, he has been a Professor at Sun Yat-sen University. Since 2019, he has also been an Adjunct Professor with the Southern Marine Science and Engineering Guangdong Laboratory (Zhuhai). His research interests are in the general areas of wireless communications and signal processing.
\end{IEEEbiography}

\vfill
\end{document}